\documentclass[12pt]{article}
\usepackage{amsfonts}
\usepackage{amssymb}
\usepackage{graphicx}
\usepackage[english]{babel}
\renewcommand{\theequation}{\arabic{section}.\arabic{equation}}

\newcommand{\field}[1]{\ensuremath{\mathbb{#1}}}

\def\be{\begin{equation}}
\def\ee{\end{equation}}
\def\bea{\begin{eqnarray}}
\def\eea{\end{eqnarray}}

\newcommand{\tg}{\tilde{\gamma}}

\begin{document}

\input epsf

\begin{flushright}OHSTPY-HEP-T-04-006\\ hep-th/0406103
\end{flushright}
\vspace{20mm}
\begin{center} {\LARGE  3-charge geometries and their CFT duals
}
\\
\vspace{20mm} {\bf  Stefano Giusto, Samir D. Mathur and
Ashish Saxena}\\
\vspace{4mm} Department of Physics,\\ The Ohio State
University,\\ Columbus, OH 43210, USA\\
\vspace{4mm}
\end{center}
\vspace{10mm}
\begin{abstract}

We consider two families of D1-D5-P states and find their gravity
duals. In each case the geometries are
found to `cap off' smoothly near $r=0$; thus there are no horizons or
closed timelike curves. These constructions support the general
conjecture that the interior of black holes is nontrivial all the way
up to the horizon.

\end{abstract}
\thispagestyle{empty}
\newpage
\setcounter{page}{1}
\section{Introduction} \setcounter{equation}{0}

The traditional picture of a black hole has a horizon, a central
singularity, and essentially `empty space' in between.
This picture leads to contradictions with quantum mechanics --
Hawking radiation leads to a loss of unitarity
\cite{hawking}. More recently a different picture of the black hole
interior has been suggested, where the information of the state of
the hole
is distributed throughout the interior of the horizon, creating a
`fuzzball' \cite{emission}. While the general state of a Schwarzschild
hole is
expected to be
very non-classical inside the horizon, we expect that for extremal
holes we can find appropriately selected states that will be
represented by classical solutions. In \cite{lm4} it was found that the
generic state  of the 2-charge
extremal D1-D5 system could be understood by studying
classical solutions of supergravity, and in
\cite{lunin},\cite{mss},\cite{gms}
classical solutions were constructed for specific families of
3-charge extremal
D1-D5-P states. The traditional picture of the 3-charge extremal hole
is pictured
in Fig.1(a). But the 3-charge geometries constructed in \cite{mss, gms}
were of the form Fig.1(b); the throat `caps off' without any  horizon
or singularity.
(All 2-charge extremal states have a geometry that caps off
like Fig.1(b); the  `naive
geometry' in this case has a zero area horizon coinciding with the
singularity at $r=0$.)

In this paper we pursue this program further, by finding further sets
of 3-charge extremal CFT states and their dual geometries. One can
write down a large class of 3-charge extremal solutions of classical
supergravity, and these will in general have
pathologies. But a basic tenet of our conjecture is that the geometries
that are dual to actual microstates of the 3-charge CFT will be regular
solutions with no horizons, singularities or closed timelike curves.
The solutions in \cite{mss, gms} were smooth, and we will find that the
solutions we now construct will also be free of horizons and closed 
timelike curves
(though most will have an orbifold singularity along certain curves).
Thus our solutions will be like Fig.1(b) rather than 1(a), and will
    lend support to the general `fuzzball' picture of the black hole
interior.\footnote{Additional evidence for this picture comes from a
study of
the nonzero size of supertubes \cite{supertubes,bk,bena,marolf}.}

In more detail, we do the following:

\medskip

(a) In \cite{bal, mm} a family of  D1-D5 geometries was obtained, by
taking extremal limits in the general family of rotating 3-charge
solutions
constructed in \cite{cy}. This family is labelled by a parameter
$0<\gamma\le 1$. The geometries have no horizons and the only
singularity is an orbifold singularity along an $S^1$ in the noncompact 
directions.  The
corresponding CFT duals can be identified \cite{bal, mm, lm4}. The
orbifold singularity
vanishes for the special case
$\gamma=1$.

         If we perform a spectral flow on the left sector of the CFT then
from a 2-charge D1-D5 state we get a 3-charge D1-D5-P state.  In
\cite{gms} we
found the geometries for the 3-charge states that are related by
spectral flow to the 2-charge state with $\gamma=1$. Here we extend
this
computation to find the geometries for  3-charge states starting from
2-charge states for arbitrary $\gamma$. It is straightforward
to identify the corresponding CFT duals.

\smallskip

(b) Given a D1-D5-P state we can use dualities to interchange any two
of its charges. This leads to a new geometry which must also
represent a
true state of the 3-charge system, since these dualities are exact
symmetries of the theory.   We construct these  geometries obtained
by S,T
dualities. These geometries turn out to have  orbifold singularities 
along two nonintersecting $S^1$ curves.
These geometries and the ones obtained in (a) all fall
into a general class that we identify; they are rotating extremal
solutions with
parameters such that there are no horizons and no closed
timelike curves.

\smallskip

(c) It is not immediately obvious what the  CFT states corresponding
to the geometries in (b) are.  To find information about the state,
we study the infall of a quantum down the `throat' of this geometry,
and study
the travel time
$\Delta t_{SUGRA}$ for a complete `bounce'. It was found in
\cite{lm4} that this bounce time exactly equalled the time $\Delta
t_{CFT}$ for excitations to travel around the corresponding
`effective string' in
the CFT.  By computing $\Delta t_{SUGRA}$ for the geometries we
find the length of each component of the effective string,  and thus
identify the CFT state.

\smallskip

(d)  We observe that the result found in (c) supports  the  picture of
``spacetime bits' arrived at in
\cite{lm4}: Under duality the number of components of the effective
string remains the same, though the total winding number of the
effective string changes. We also observe that the travel time in the
geometry and in the CFT
are related by a redshift factor $\eta$ which relates the time
coordinate at infinity to the time coordinate
in the AdS region ($\eta$ becomes unity if the momentum charge P
vanishes).

\smallskip

\begin{figure}[ht]
\hspace{0in}
\includegraphics{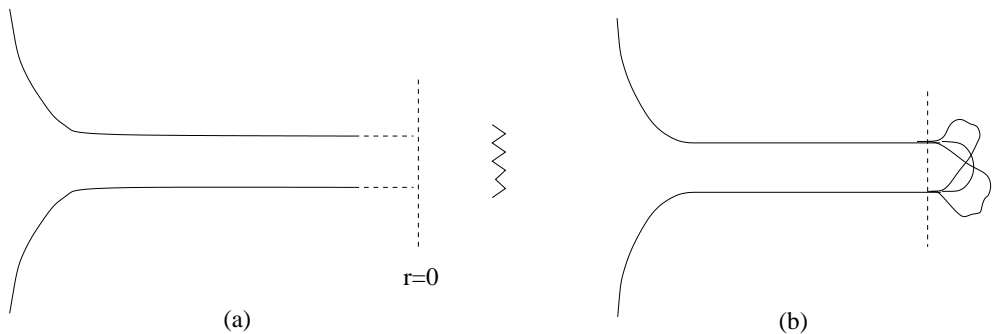}
\caption{\small{(a) Naive geometry of 3-charge D1-D5-P; there is a
horizon at $r=0$ and a singularity past
the horizon. (b) Expected geometries for D1-D5-P; the area at the
dashed line will give
${A\over 4G}=2\pi\sqrt{n_1n_5n_p}$.}}
\label{fig}
\end{figure}

\section{D1-D5-P states from spectral flow of D1-D5
states}\setcounter{equation}{0}

\subsection{The D1-D5 CFT}

We take IIB string theory compactified to $M_{4,1}\times S^1\times
T^4$.  Let $y$ be the coordinate along $S^1$ with
\be
0\le y<2\pi R
\ee
The $T^4$ is described by 4 coordinates $z_1, z_2, z_3, z_4$, and
the noncompact space is spanned by $t, x_1, x_2, x_3, x_4$.  We
wrap
$n_1$ D1 branes on $S^1$, and
$n_5$ D5 branes on
$S^1\times T^4$. Let $N=n_1n_5$. The bound state of these branes
is described by a 1+1 dimensional sigma model, with base space
$(y,t)$ and target space a deformation of the orbifold
$(T^4)^N/S_N$ (the symmetric product of $N$ copies of $T^4$).  The
CFT has ${\cal N}=4$ supersymmetry, and a moduli space which
preserves this supersymmetry. It is conjectured that in this
moduli space we have an `orbifold point' where the target space is
just the orbifold
$(T^4)^N/S_N$ \cite{sw}.

The rotational symmetry of the noncompact directions $x_1\dots
x_4$ gives a symmetry
$so(4)\approx su(2)_L\times su(2)_R$, which is the R symmetry group of
the CFT.

The CFT with target space just one copy of $T^4$ is described by 4
real bosons
$X^1$,  $X^2$,  $X^3$,  $X^4$ (which arise from the 4 directions $z_1,
z_2,
z_3, z_4$), 4 real left moving fermions $\psi^1, \psi^2, \psi^3,
\psi^4$ and 4 right moving fermions $\bar\psi^1, \bar\psi^2,
\bar\psi^3, \bar\psi^4$. The central charge is $c=6$.  The complete
theory with target space $(T^4)^N/S_N$ has $N$ copies of this
$c=6$ CFT, with states that are symmetrized between the $N$
copies. The orbifolding also generates `twist' sectors, which are
created by twist operators $\sigma_k$. A detailed construction
of the twist operators is given in  \cite{lm1, lm2}, but we summarise
here the properties that will be relevant to us.

The twist operator of order $k$ links together $k$ copies of the $c=6$
CFT so that the $X^i, \psi^i$ act as free fields living on a circle
of length $kL$ ($L$ is the length of the spatial circle of the CFT).
Let us first discuss the NS sector.
The left fermions
$\psi^i$ carry spin
${1\over 2}$ under the  $su(2)_L$ and the right fermions
$\bar\psi^i$ carry  spin
${1\over 2}$ under the  $su(2)_R$.  The `charge' of a state is
given by the quantum numbers $(j, \bar j)=(j^3_L, j^3_R)$. Adding a
suitable charge to the twist
operator we get a chiral primary
\be
\sigma_k^{--}: ~~h=j={k-1\over 2}, ~~\bar h=\bar j={k-1\over 2}
\ee
We can act on this chiral primary with $J^+_{-1}$ to get another chiral
primary
\be
\sigma^{+-}_k\equiv J^+_{-1}\sigma^{--}_k:~~h=j={k+1\over 2}, ~~\bar
h=\bar j={k-1\over 2}
\ee
Similarly we also get
\be
\sigma^{-+}_k\equiv  \bar J^+_{-1}\sigma^{--}_k:~~h=j={k-1\over 2},
~~\bar h=\bar j={k+1\over 2}
\ee
\be
\sigma^{++}_k\equiv J^+_{-1}\bar J^+_{-1}\sigma^{--}_k:~~h=j={k+1\over
2}, ~~\bar h=\bar j={k+1\over 2}
\ee
(We can get additional chiral primaries by applying for example
$\psi^+_{-{1\over 2}}$ (which increase $h$ and $j$ by ${1\over 2}$,
but we will not need such states in this paper.))

\subsubsection{ A subclass of states}

In the NS sector we can start with the NS vacuum
\be
|0\rangle_{NS}:~~h=j=0,~ \bar h=\bar j=0
\ee
and act with $\sigma^{\pm\pm}_k$ to generate chiral primaries. Consider
the subclass of states
\be
(\sigma^{--}_k)^{N\over k}|0\rangle_{NS}:~~h=j={N\over k}{(k-1)\over
2}, ~~\bar h=\bar j={N\over k}{(k-1)\over 2}
\label{one}
\ee
All copies of the CFT are linked into `long circles' which are all of
the same length $kL$, and the spin orientation
(given by the choice $(--)$ for $\sigma$) is also the same for each
circle. We therefore expect that the corresponding states exhibit some
symmetry; it will turn out that their gravity duals have axial symmetry
around two circles $\psi, \phi$.  Similarly we have the
states
\be
(\sigma^{+-}_k)^{N\over k}|0\rangle_{NS}:~~h=j={N\over k}{(k+1)\over
2}, ~~\bar h=\bar j={N\over k}{(k-1)\over 2}
\label{s+-}
\ee
\be
(\sigma^{-+}_k)^{N\over k}|0\rangle_{NS}:~~h=j={N\over k}{(k-1)\over
2}, ~~\bar h=\bar j={N\over k}{(k+1)\over 2}
\ee
\be
(\sigma^{++}_k)^{N\over k}|0\rangle_{NS}:~~h=j={N\over k}{(k+1)\over
2}, ~~\bar h=\bar j={N\over k}{(k+1)\over 2}
\ee

\subsubsection{ Spectral flow}

        The NS sector
states can be mapped to R sector states by `spectral flow'
\cite{spectral}, under which the conformal dimensions and
charges change as
\bea
h'&=&h-\alpha q + \alpha^2{c\over 24}\\
q'&=&q-\alpha{c\over 12}
\label{qone}
\eea
Setting $\alpha=1$ gives the flow from the NS sector to the R
sector, and we can see that under this flow chiral primaries
of the NS sector
(which have $h=q$) map to Ramond ground states with $h={c\over
24}$.

The field theory on the D1-D5 branes system is in the R sector.
This follows from the fact that the branes are solitons of
the gravity theory, and the fermions on the branes are induced
from fermions on the bulk. The latter are periodic around the
$S^1$; choosing antiperiodic boundary conditions would give a
nonvanishing vacuum energy and disallow the flat space solution
that we have assumed at infinity.

If we set $\alpha=2$ in (\ref{qone}) then we return to the NS
sector, and setting $\alpha=3$ brings us again to the R sector.
More generally, the choice
\be
\alpha=2n+1\,, ~~~ n\in\mathbb{Z}
\label{qtwo}
\ee
brings us to the R sector.

\subsubsection{The states we consider}

Suppose we start with a chiral primary in the NS sector. Perform a
spectral flow (\ref{qtwo}) on the right movers
with $\alpha=1$; this brings us to an R ground state for the right
movers. Perform a spectral flow with
$\alpha=2n+1$ on the left movers. This brings us to the R sector but
not in general to an R ground state.
The state thus has a momentum charge
\be
n_p=h-\bar h
\ee
Applying this procedure to the state (\ref{one}) we get an R sector
state
\be
[(\sigma^{--}_k)^{N\over k}|0\rangle_{NS}]_{\alpha_L=2n+1,
\alpha_R=1}\equiv |\Psi^{--}(k,n)\rangle
\label{two}
\ee
with
\be
h=N\Bigl(n^2+{n\over k}+{1\over 4}\Bigr), ~~j=-{N\over
2}\Bigl(2n+{1\over k}\Bigr), ~~\bar
h={N\over 4}, ~~\bar j=-{N\over 2k}
\label{twop}
\ee
and
\be
n_p=h-\bar h=N\,n\Bigl(n+{1\over k}\Bigr)
\ee

\subsubsection{Explicit representations of the states}

Let us construct explicitly the above CFT states. Consider one copy
of the $c=6$ CFT, in the R sector.  The  fermions have modes
$\psi^i_m$. The 4 real fermions can be
grouped into 2 complex fermions  $\psi^+, \psi^-$ which form  a
representation of $su(2)$.  ($\psi^+$ has $j={1\over 2}$ and
$\psi^-$ has $j=-{1\over 2}$.)  The anti-commutation relations
are
\be
\{(\psi^+)^*_m, \psi^+_p\}=\delta_{m+p,0}, ~~~\{(\psi^-)^*_m,
\psi^-_p\}=\delta_{m+p,0}
\ee
         The
$su(2)$ currents are
\be
J^+_m=(\psi^-)^*_{m-p}\psi^+_p,
~~~J^-_m=(\psi^+)^*_{m-p}\psi^-_p, ~~~J^3_m={1\over 2}
[(\psi^-)^*_{m-p}\psi^-_p-(\psi^+)^*_{m-p}\psi^+_p]
\ee
In the full theory with $n_1n_5$ copies of the
$c=6$ CFT  the currents are the sum of the currents in
the individual copies
\be
J^{a,total}_{ n}=(J^a_{n})_1+\dots (J^a_{n})_{n_1n_5}
\ee

First consider the state $|\Psi^{--}(k,0)\rangle$ ((\ref{two}) for
$n=0$). This gives a D1-D5 state with momentum charge zero
\be
[(\sigma^{--}_k)^{N\over k}|0\rangle_{NS}]_{\alpha_L=1,
\alpha_R=1}:~~h={N\over 4}, ~~j=-{N\over 2k},
~~\bar h={N\over 4}, ~~\bar j=-{N\over 2k}, ~~n_p=0
\label{2charge}
\ee
Each set of $k$ copies of the $c=6$ CFT which are joined together by
$\sigma_k^{--}$ behave like one
copy of the $c=6$ CFT but on a circle of length $kL$.

\smallskip

{\bf Note:} We will call each such set of linked copies a {\it component
string}.

\smallskip

Thus in the
presence of a twist operator of order  $k$ we can apply fractional
modes of currents
\be
J^a_{-{m\over k}}, ~~\bar J^a_{-{m\over k}}
\ee
Since we are in the R sector the fermions have fractional modes
$\psi^\pm_{-{m\over k}}$. Apart from this fractionation
the situation is identical to the one studied in \cite{gms} where we
had no twist ($k=1$) and we applied currents to
find the states arising after spectral flow. In the present case the
lowest dimension current operator that we can apply to lower charge is
$J^-_{-{2\over k}}$ to (\ref{2charge}); this is equivalent to applying
$(\psi^-)^*_{-{1\over k}}\psi^-_{-{1\over k}}$. The next operator we
can apply is  $J^-_{-{4\over k}}$, and so on. The orbifold CFT requires
that
the total momentum on each component string be an integer -- we will
discuss this issue
(and possible exceptions)  in more detail
in section \ref{cftdualsSec}.
Thus we can apply
\be
J^-_{-{2(k-1)\over k}}\dots J^-_{-{4\over k}}J^-_{-{2\over k}}
\label{first}
\ee
which adds dimension and charge
\be
\Delta h=k-1, ~~\Delta j=-(k-1)
\ee
or we can apply
\be
J^-_{-2}J^-_{-{2(k-1)\over k}}\dots J^-_{-{4\over k}}J^-_{-{2\over k}}
\label{second}
\ee
which adds dimension and charge
\be
\Delta h=k+1, ~~\Delta j=-k
\ee
To get states with a high symmetry we apply the same set of current
operators to all the ${N\over k}$ twist operators in the state.
Applying (\ref{first}) we get a state
\be
(\prod [J^-_{-{2(k-1)\over k}}\dots J^-_{-{4\over k}}J^-_{-{2\over
k}}])|\Psi^{--}(k,0)\rangle
\label{fstate}
\ee
where the product runs over the ${N\over k}$ connected components of
the CFT created by the twists. This state has
dimensions and charges
\be
h={N\over 4}+{N\over k}(k-1), ~~j=-{N\over 2k}-{N\over k}(k-1), ~~\bar
h={N\over 4}, ~~\bar j=-{N\over 2k}
\ee
and
\be
n_p={N\over k}(k-1)
\ee
Applying (\ref{second}) instead to each of the component strings we get
the state
\be
(\prod [J^-_{-2}J^-_{-{2(k-1)\over k}}\dots J^-_{-{4\over
k}}J^-_{-{2\over
k}}])|\Psi^{--}(k,0)\rangle
\label{sstatepre}
\ee
        with dimensions and charges
\be
h={N\over 4}+{N\over k}(k+1), ~~j=-{N\over 2k}-N, ~~\bar h={N\over 4},
~~\bar j=-{N\over 2k}
\ee
\be
n_p={N\over k}(k+1)
\ee

We now observe that the state (\ref{sstatepre}) has the correct
dimensions
and charges to be the member of the spectral flow
family (\ref{two}) with $n=1$. Similarly, the state (\ref{fstate})
can be identified with the state obtained by spectral flow,
with $n=1$, from the state $(\sigma^{+-}_k)^{N\over k}|0\rangle_{NS}$ in
eq. (\ref{s+-}).

We can apply further sets of currents to get states with spectral flow
by $n>1$ units
\be
(\prod [J^-_{-2n}\dots J^-_{-{2\over
k}}])|\Psi^{--}(k,0)\rangle
\label{sstate}
\ee
     These states have
\be
h={N\over 4}+{N\over k} n (n k+1), ~~j=-{N\over 2k}-n N, ~~\bar
h={N\over 4},
~~\bar j=-{N\over 2k}
\ee
\be
n_p={N\over k}n (n k+1)
\ee
We can similarly get those with $n<0$;  latter are obtained by
applying modes of $J^+$ instead of $J^-$.

\subsection{Gravity duals}

\subsubsection{Duals of 2-charge states}

In \cite{cy} a  set of  D1-D5-P solutions was given. The  solutions had
        axial symmetry along
two circles $\psi, \phi$, and angular momenta $J_\psi, J_\phi$. In
\cite{bal, mm} an extremal limit was obtained for solutions with $P=0$,
getting the geometries
\bea
ds^2 & = & -\frac{1}{h} (dt^2-dy^2) + h f \left( \frac{dr^2}{r^2 +
a^2\,\gamma^2} + d\theta^2 \right)
\nonumber \\
         &+& h \Bigl( r^2 +
\frac{a^2\,\gamma^2\,Q_1 Q_5\,\cos^2\theta}{h^2 f^2} \Bigr)
\cos^2\theta d\psi^2  \nonumber \\
&+& h\Bigl( r^2 + a^2\,\gamma^2 -
\frac{a^2\,\gamma^2\,Q_1 Q_5 \,\sin^2\theta}{h^{2} f^{2} }
\Bigr) \sin^2\theta d\phi^2  \nonumber \\
&-& \frac{2a\,\gamma\,\sqrt{Q_{1}Q_{5}} }{hf}
(\cos^2\theta \,dy\,d\psi + \sin^2\theta \,dt\,d\phi)+\sqrt{H_1\over
H_5}\sum_{i=1}^4 dx_i^2
\label{mm}
\eea
where
\bea
&&a={\sqrt{Q_1 Q_5}\over R}\,,\quad
f=r^2+a^2\,\gamma^2\,\cos^2\theta\nonumber\\
&&H_1 =1+{Q_1\over f}\,,\quad H_5 =1+{Q_5\over f}\,,\quad  h =
\sqrt{H_1 H_5}
\eea
These metrics have angular momenta
\be
J_\psi=-{\bar j}+j = 0 \,,\quad J_\phi = -{\bar j}-j= \gamma\,n_1 n_5
\ee
with $n_1$ and $n_5$ the numbers of D1 and D5 branes.  $j, \bar j$ are
the angular momenta in the
$L,R$ factors of the $so(4)\approx su(2)_L\times su(2)_R$ describing
the angular directions. For
\be
\gamma={1\over k}, ~~k=1,2\dots
\ee
we obtain geometries that are the duals  of the states
$|\Psi^{--}(k,0)\rangle$ (eq. (\ref{2charge})) \cite{bal, mm, lm4}.

\subsubsection{Duals of 3-charge states obtained by spectral flow of
2-charge states}

We would now like to find the duals of the 3-charge states obtained by
        spectral
flow of the above 2-charge states. We again start from the 3-charge
nonextremal solutions and take an extremal limit,
keeping the charges and angular momenta at the values given by the CFT
state. For the case where all twists were trivial
($k=1$ for all twist operators) this procedure was carried out in
\cite{gms}. The starting nonextremal solution was derived
in \cite{gms} by starting with the neutral rotating hole in 4+1
dimensions, and applying a sequence of boosts and dualities.
This solution is reproduced in Appendix \ref{app1}. Taking the limit
for $P\ne
0$  needs some care, but
the procedure was described in detail in \cite{gms} and needs no
changes for the more general case here.

We fix the values of the angular momenta and the momentum charge $n_p$
to the values we desire,
and then take the extremal limit. For general values of these
parameters, we obtain the solution
given in Appendix \ref{app1}. These solutions have pathologies in
general; for
example they can have closed timelike curves. But we
must select only those that correspond to microstates of the 3-charge
system, and these we expect to be free of pathologies.

We set the charges and momentum to equal those of the states
(\ref{sstate})
\be
J_\psi = -n\, n_1 n_5  \,,\quad J_\phi = (n+\gamma)\,n_1 n_5, ~~
n_p = n(n+\gamma)\,n_1 n_5
\ee
The metric is written naturally in terms of the dimensionful quantities
\bea
&&\tg_1 = {4\,G^{(5)}\over \pi
\sqrt{Q_1 Q_5}}\,J_\psi \,, ~~~~\tg_2= {4\,G^{(5)}\over \pi \sqrt{Q_1
Q_5}}\,J_\phi\nonumber\\
&&Q_p = {4\,G^{(5)}\over \pi R}\,n_p\,, ~~~~ Q_1 = {(2\pi)^4 g
\alpha'^3\over V}\,n_1\,,~~~~ Q_5= g\alpha'\, n_5
\eea
where $G^{(5)}$ is the 5-D Newton's constant
\be
G^{(5)}= {G^{(10)}\over V (2\pi R)}={4 \pi^5 g^2 \alpha'^4\over V R}
\ee
$V$ is the volume of $T^4$, $R$ the radius of the $S^1$ and $g$ the
string coupling.
We find
\be
\tg_1={\sqrt{Q_1 Q_5}\over R}\,{J_\psi\over n_1 n_5}\equiv{\sqrt{Q_1
Q_5}\over R}\,\gamma_1 \,,\quad
\tg_2={\sqrt{Q_1 Q_5}\over R}\,{J_\phi\over n_1 n_5}\equiv{\sqrt{Q_1
Q_5}\over R}\,\gamma_2
\label{gamma1}
\ee
and thus, for the duals of the states (\ref{sstate}),
\be
\tg_1 =  -{\sqrt{Q_1 Q_5}\over R}\,n\,,\quad \tg_2 =
{\sqrt{Q_1 Q_5}\over R}\,(n+\gamma),\quad Q_p  ={Q_1 Q_5\over R^2}\,
n(n+\gamma)
\label{gamma2}
\ee
We observe that
\be
Q_p=
-\tg_1 \,\tg_2
\label{condition}
\ee
We will see that it is this relation that selects, from the class of
all axisymmetric solutions,  the geometries that
are free of pathologies.\footnote{We certainly have states of the
system with $Q_p\ne 0$ and small or vanishing angular momenta, but as
seen in the 2-charge case \cite{lm4} such states will break axial
symmetry, and thus not be in the class that we are considering at
present.}
     Using
(\ref{condition}) to simplify the extremal solution, we get
\bea
\label{extremalmetric}
ds^2 & = & -\frac{1}{h} (dt^2-dy^2) + \frac{Q_{p}}{h
f}\left(dt-dy\right)^{2}+ h f \left( \frac{dr^2}{r^2 +
(\tg_1+\tg_2)^2\eta} + d\theta^2
\right)\nonumber \\
         &+& h \Bigl( r^2 + \tg_1\,(\tg_1+\tg_2)\,\eta -
\frac{Q_1 Q_5\,(\tg_1^2-\tg_2^2)\,\eta\,\cos^2\theta}{h^2 f^2}
\Bigr)
\cos^2\theta d\psi^2  \nonumber \\
&+& h\Bigl( r^2 + \tg_2\,(\tg_1+\tg_2)\,\eta +
\frac{Q_1 Q_5\,(\tg_1^2-\tg_2^2)\,\eta\,\sin^2\theta}{h^{2} f^{2}
}
\Bigr) \sin^2\theta d\phi^2  \nonumber \\
&+& \frac{Q_p\,(\tg_1+\tg_2)^2\,\eta^2}{h f}
\left( \cos^2\theta d\psi + \sin^2\theta d\phi \right)^{2} \nonumber\\
&-& \frac{2 \sqrt{Q_{1}Q_{5}} }{hf}
\left(\tg_1 \cos^2\theta d\psi + \tg_2 \sin^2\theta d\phi\right)
(dt-dy)
\nonumber \\
&-& \frac{2 \sqrt{Q_1 Q_5}\,(\tg_1+\tg_2)\,\eta}{h f}
\left( \cos^2\theta d\psi + \sin^2\theta d\phi \right) dy +
\sqrt{H_1\over H_5}\sum_{i=1}^4 dx_i^2
\eea
\bea
C_{2} &=& - \frac{\sqrt{Q_{1} Q_{5}} \cos^2\theta}{H_1 f }
\left(\tg_2\,dt + \tg_1\,dy \right)
\wedge d\psi
- \frac{\sqrt{Q_{1}Q_{5} } \sin^2\theta }{H_1 f }
\left(\tg_1\,dt + \tg_2\,dy \right) \wedge d\phi
\nonumber \\
&+& \frac{(\tg_1+\tg_2)\,\eta\,Q_{p} }{\sqrt{Q_{1}Q_{5}}H_1 f }
\left( Q_{1}\,dt + Q_{5}\,dy \right)
\wedge \left( \cos^2\theta d\psi  + \sin^2\theta d\phi \right)
\nonumber \\
&-& \frac{ Q_{1} }{H_1 f} dt \wedge dy - \frac{ Q_{5}\cos^2\theta
}{H_1 f} \left( r^{2} +
\tg_2 (\tg_1+\tg_2)\,\eta + Q_{1} \right)  d\psi \wedge d\phi
\eea
\be
e^{2\Phi}= \frac{H_{1}}{H_{5}}
\ee
where
\bea
&&\!\!\!\!\!\!\!\!\!\!\!\!\eta = {Q_1 Q_5\over Q_1 Q_5 + Q_1 Q_p + Q_5
Q_p}
\nonumber\\
&&\!\!\!\!\!\!\!\!\!\!\!\!f = r^2+ (\tg_1+\tg_2)\,\eta\,
\bigl(\tg_1\, \sin^2\theta + \tg_2\,\cos^2\theta\bigr) \nonumber\\
&&\!\!\!\!\!\!\!\!\!\!\!\!H_{1} = 1+
\frac{Q_{1}}{f}\,,\quad H_{5} = 1+ \frac{Q_{5}}{f}\,,\quad h =
\sqrt{H_{1} H_{5}}
\label{deffh}
\eea

\section{Obtaining new solutions by S,T
dualities}\setcounter{equation}{0}

As mentioned in the introduction, we are interested in making
geometries that are dual to actual bound states of the D1-D5-P system,
and not just formal solutions of supergravity carrying D1,D5,P charges.
In the previous section we started with known states in the CFT and
found their gravity duals by looking for solutions with the same
symmetries and quantum numbers.
In this section we
make
D1-D5-P solutions by a different method: We start with a D1-D5-P
geometry that we have already constructed and perform
S,T dualities to permute the charges. Since these dualities are exact
symmetries of the theory, we know that the resulting geometry
represents a true microstate. But it will not be immediately obvious
what the dual CFT state is. We will identify the CFT state later, by
analyzing the properties of the supergravity solution, and find that
the change of CFT state under these S,T dualities provides insight into
the
AdS/CFT duality map.

The metric (\ref{extremalmetric}) is invariant under the interchange  of
$Q_1 , Q_5$, so the only non-trivial duality
is the one which interchanges the momentum charge with, say, the D1
charge
\be
\left(\begin{array}{c} P \\ D1 \\ D5 \end{array} \right)
\stackrel{S}{\longrightarrow}
\left(\begin{array}{c} P \\ F1 \\ NS5 \end{array}
\right)\stackrel{T_{y},T_{z_{1}}}{\longrightarrow}
\left(\begin{array}{c} F1 \\ P \\ NS5 \end{array} \right)
\stackrel{S}{\longrightarrow}
\left(\begin{array}{c} D1 \\ P \\ D5 \end{array} \right)
\label{chain}
\ee
The metric which results from this chain of dualities, and the
corresponding dilaton are
\bea
ds^2 & = & -\frac{1}{\tilde h} (dt^2-dy^2) + \frac{Q_{1}}{{\tilde h}
f}\left(dt-dy\right)^{2} + {\tilde h} f \left( \frac{dr^2}{r^2 +
(\tg_1+\tg_2)^2\eta}+ d\theta^2 \right)
\nonumber \\
         &+& {\tilde h}\Bigl( r^2 + \tg_1\,(\tg_1+\tg_2)\,\eta -
\frac{Q_p Q_5\,(\tg_1^2-\tg_2^2)\,\eta\,\cos^2\theta}{{\tilde
h}^2 f^2} \Bigr)
\cos^2\theta d\psi^2  \nonumber \\
&+& {\tilde h} \Bigl( r^2 + \tg_2\,(\tg_1+\tg_2)\,\eta +
\frac{Q_p Q_5\,(\tg_1^2-\tg_2^2)\,\eta\,\sin^2\theta}{{\tilde
h}^2 f^{2} }
\Bigr) \sin^2\theta d\phi^2  \nonumber \\
&+& \frac{Q_p\,(\tg_1+\tg_2)^2\,\eta^2}{{\tilde h} f}
\left( \cos^2\theta d\psi + \sin^2\theta d\phi \right)^{2} \nonumber\\
&-& \frac{2 \sqrt{Q_{1}Q_{5}} }{{\tilde h} f}
\left(\tg_1 \cos^2\theta d\psi + \tg_2 \sin^2\theta d\phi\right)
(dt-dy)
\nonumber \\
&-& \frac{2 Q_p \,(\tg_1+\tg_2)\,\eta}{{\tilde h}
f}\sqrt{Q_5\over Q_1}
\left( \cos^2\theta d\psi + \sin^2\theta d\phi \right) dy +
\sqrt{H_p\over H_5}\sum_{i=1}^4 dx_i^2\nonumber
\label{pthree}
\eea
\bea
C_{2} &=& -\frac{\sqrt{Q_{1} Q_{5}} \cos^2\theta}{H_p f }
\left(\tg_2
dt + \tg_1 dy \right) \wedge d\psi -
      \frac{\sqrt{Q_{1}Q_{5} } \sin^2\theta}{H_p f} \left(\tg_1 dt +
\tg_2  dy \right) \wedge d\phi \nonumber \\
&+& \frac{(\tg_1 + \tg_2)\eta }{H_p f }
\sqrt{\frac{Q_{1}}{Q_{5}}}\left( Q_{p}
dt + Q_{5} dy \right) \wedge \left( \cos^2\theta d\psi  + \sin^2\theta
d\phi \right) \nonumber \\
&-& \frac{ Q_{p} }{H_p f} dt \wedge dy - \frac{ Q_{5}\cos^2\theta
}{H_p f} \left
( r^{2} + \tg_2 (\tg_1+\tg_2) \eta + Q_{p} \right)  d\psi
\wedge d\phi
\nonumber
\eea
\be
e^{2\Phi}= \frac{H_{p}}{H_{5}}
\ee
where
\be
H_{p} = 1+ \frac{Q_{p}}{f}\,,\quad {\tilde h}= \sqrt{H_p\, H_5}
\ee

The solution above is again of the general form (\ref{extremalmetric})
but with different parameters\footnote{The quantities $Q_i$ have units
of $(length)^2$, so we have to be careful about
the meaning of (\ref{sts}). If we start with the F-NS5-P system and do
$T_y$ to get P-NS5-F then we get
${Q'_1\over l_s'^2}={Q_p\over l_s^2}, ~{Q'_p\over l_s'^2}={Q_1\over
l_s^2}$ etc., where $l_s$ is the string length. Starting with D1-D5-P
and applying $S T_{y} T_{z_{1}} S$ to get P-D5-D1 we get ${Q'_1\over 
l_d'^2}={Q_p\over
l_d^2}, ~{Q'_p\over l_d'^2}={Q_1\over l_d^2}$ where $l_d=g^{1\over 2}
l_s$ is the D-string length. Since the classical geometry is unchanged
by an overall rescaling,
we can write (\ref{sts}).}
\bea
&&Q'_1 = Q_p\,,\quad Q'_5 = Q_5\,,\quad Q'_p = Q_1\nonumber\\
&&\tg'_1=\sqrt{Q_1\over
Q_p}\,\tg_1\,,\quad\tg'_2=\sqrt{Q_1\over Q_p}\,\tg_2
\label{sts}
\eea

\section{Conical defect angles}\setcounter{equation}{0}

The geometries (\ref{extremalmetric}) possess, generically,  an 
orbifold singularity
  along a circle in the noncompact space directions, just like the
subset of 2-charge metrics (\ref{mm}). Let us recall the physics of
these singularities,
and then study the `conical defect' angle (created in the AdS part of 
the geometry by the
orbifolding) for the metrics constructed in the
above sections.

\subsection{The physics of conical defects}

Let us recall the reason why we get conical defects in 2-charge
geometries. In \cite{lm4} a method was developed to compute the gravity
duals for {\it all} states of the 2-charge system (not just the
subclass giving the geometries (\ref{mm})).
By S,T dualities we map the D1-D5 system to the FP system, which has a
fundamental
string (F) wrapped on $S^1$ carrying momentum (P) along $S^1$. The
bound state of the FP system
has the strands of the F string all joined up into one `multiwound'
string, and all the momentum is
carried as travelling waves on the string.
Metrics for the vibrating string were constructed,
and dualized back to get D1-D5 geometries. The general geometry was
thus parametrized by the vibration profile
$\vec F(v)$ of the F string.

The detailed map between D1-D5 states and D1-D5 geometries is found in
the following way.
The vibration  on the F string is written in terms of harmonics. If we
have a quantum of the $k$th harmonic
on the string then we get a twist operator $\sigma_k$ acting on the NS
vacuum in the D1-D5 CFT.
The polarization of the vibration is given by an index $i=1\dots 4$
labeling the four noncompact directions.
The $so(4)$ symmetry group of the angular directions  is $\approx
su(2)_L\times su(2)_R$, and writing the vector
index $i$ in terms of the representation $({1\over 2}, {1\over 2})$ of
$su(2)_L\times su(2)_R$ we get the choice of
superscripts of the twist operator $\sigma_k^{\pm, \pm}$. The
collection of all twist operators (arising from all
quanta of vibration on the F string) give an NS state, which is
spectral flowed to get an R sector state in the D1-D5 CFT.
The geometry for this state is known, since it is obtained by $S,T$
dualities from the geometry created by the vibrating F string.

The strands of the F string wrap the $y$ direction, but under the
vibration they carry they separate out
from each other, and spread out over a simple closed curve  in the
transverse space $x^1\dots x^4$. It appears at first that there would
be a singularity in the FP and D1-D5 geometries at this curve, but it
was found in \cite{lm4} that all waves
reflect trivially off this singularity. The reason for this was
explained in \cite{lmm} where it was found that for the D1-D5
geometries the singularity was just a {\it coordinate} singularity
similar to that at the origin of a KK monopole; we have
a `KK monopole tube'  (KK monopole $\times S^1$) centered at the above
curve. As long as this curve does not self-intersect (it generically
does not self-intersect since it is just an $S^1$ in $\field{R}^4$) the
2-charge D1-D5 geometry is completely smooth, with no horizon or
singularity.

The generic solution has no particular symmetry. If we look at
solutions that have axial symmetry then there are
very few possibilities. We must let the F string swing in a uniform
helix in the covering space of the $y$ circle; let
this helix have $k$ turns. The vibrations are now all in the $k$th
harmonic, so the D1-D5 CFT state is created
by $(\sigma^{--}_k)^{N\over k}$ (The choice $(--)$ says that we let the
F string swing in the $x_1-x_2$ plane; changing this plane changes the
superscripts). The KK monopole  tube now `runs over itself' $k$ times,
so all points on the $S^1$
exhibit the geometry of $k$ KK monopoles coming together. But it is
known that this generates an ALE singularity, which has a conical
defect angle of $2\pi (1-{1\over k})$.

\subsection{Singularity structure of 3-charge metrics}

We thus see that a conical defect is a `harmless singularity' -- it
arises only if we make nongeneric states by
letting the $S^1$ run over itself, and in the full quantum theory one
can imagine that quantum fluctuations
separate the intersecting strands and smooth out the singularity. It is
important though that the conical defect angle be of the form $2\pi
(1-{1\over k})$; if we find an irrational angle for instance then we
would not be able to understand how the given geometry sits in a family
of geometries that are generically smooth.
We would like to understand more precisely the nature of these
defects. In particular we would also like to know the values of the
parameters $\gamma_{1}$ and $\gamma_{2}$ for which such defects might
arise. The arguments we use below are similar to the ones
given in \cite{gms}, where more details can be found.

Around $r=0$ the 6-dimensional part of the 3-charge metric
(\ref{extremalmetric}) have the following form:
\bea
ds^2 &\approx& {hf\over (\tg_1+\tg_2)^2\,\eta}\,\Bigl(dr^2 +
r^2\,{d{\tilde y}^2\over R^2}\Bigr)\nonumber\\
\quad &+& h f\Bigl(d\theta^2 + {\tilde g}_{\psi\psi}\,
\cos^2\theta\,d{\tilde\psi}^2 +{\tilde g}_{\phi\phi}\,\sin^2\theta\,
d{\tilde \phi}^2 + 2 {\tilde g}_{\psi\phi}\,\cos^2\theta\,\sin^2\theta\,
d{\tilde \psi} d{\tilde \phi} \Bigr)\nonumber\\
\quad&+& g_{tt}\,dt^2 + 2 {\tilde g}_{t\psi}\,\cos^2\theta\,dt d{\tilde
\psi} + 2
{\tilde g}_{t\phi}\,\sin^2\theta\,dt d{\tilde \phi}
\label{r=0}
\eea
where
\be
\gamma=|\gamma_1 +\gamma_2|
\ee
\be
{\tilde y}= \gamma\,y\,,\quad {\tilde \psi} = \psi - \gamma_2\,{y\over
R}\,,\quad
{\tilde \phi}=  \phi-\gamma_1\,{y\over R}
\ee
\be
f \approx (\tg_1+\tg_2)\,\eta\,
\bigl(\tg_1\, \sin^2\theta + \tg_2\,\cos^2\theta\bigr)\nonumber\\
\ee
and ${\tilde g}_{\psi\psi}$, ${\tilde g}_{\phi\phi}$, ${\tilde
g}_{\psi\phi}$,
$g_{tt}$, ${\tilde g}_{t\psi}$, ${\tilde g}_{t\phi}$ are differentiable
functions of
$\theta$ with
\be
\tilde{g}_{\psi\psi}(\pi/2)=1\,,\quad \tilde{g}_{\phi\phi}(0)=1 
\label{polar}
\ee
The above form of the metric shows that at $r=0$ the ${\tilde y}$ cycle
shrinks. It is
important to know if, for some particular value of $\theta$, some other
cycle shrinks
at the same time. This can be understood by looking at the determinant
of the metric
restricted to the $t$, ${\tilde \psi}$ and ${\tilde \phi}$ coordinates
\be
\det g|_{t,{\tilde \psi},{\tilde \phi}}=-{(Q_1 Q_5)^2\over
R^2}{\eta^2\,\gamma^2\over hf}\,\sin^2\theta\,\cos^2\theta
\ee
This determinant only vanishes at $\theta=0,\pi/2$: At $\theta=0$ the
${\tilde \phi}$ coordinate
decouples from ${\tilde \psi}$ and $t$ and the coefficient of $d{\tilde
\phi}^2$ vanishes, i.e. the
${\tilde \phi}$ cycle shrinks at this point. At $\theta=\pi/2$ the same
happens
for the ${\tilde \psi}$ cycle. No other combination of the  $\tilde
\phi$ and $\tilde \psi$ cycles
vanishes at any other value of $\theta$.

To proceed further, we need to look at the actual values of $\gamma$,
$\gamma_1$ and $\gamma_2$.

\subsubsection{Geometries obtained by spectral flow: orbifold 
singularities}

Let us consider first
the case of the 3-charge metrics obtained by spectral flow from
the 2-charge metrics in (\ref{mm}). For them
\be
\gamma_1 = -n\,,\quad \gamma_2 = \Bigl(n+{1\over k}\,\Bigr)\,,\quad
\gamma={1\over k}\,,\quad k\in\mathbb{N}\,,\quad n\in\mathbb{Z}
\label{sf}
\ee
In this case the coordinates $\tilde y$, $\tilde \psi$ and $\tilde
\phi$ are
\be
{\tilde y}={y\over k}\,,\quad {\tilde \psi}=\psi - \Bigl(n+{1\over
k}\Bigr)\,{y\over R}\,,\quad
{\tilde \phi}=\phi + n\,{y\over R}
\label{cd}
\ee
If  ${\tilde y}/R$, $\tilde \psi$ and $\tilde \phi$ were periodic
coordinates with period $2\pi$, i.e. if
\be
\Bigl({{\tilde y}\over R},{\tilde \psi},{\tilde \phi}\Bigr)\sim
\Bigl({{\tilde y}\over R},{\tilde
\psi},{\tilde \phi}\Bigr)+2\pi(l_1,l_2,l_3)
\label{period}
\ee
with $l_1$, $l_2$, $l_3$ integers, then the metric (\ref{r=0}) would be
smooth as  can be seen from the coefficients of $dy, d\tilde{\psi}$ and 
$d\tilde{\phi}$ in Eqs. (\ref{r=0}-\ref{polar}). However, it follows 
from the definition
(\ref{cd}) and from the periodicity of the asymptotic coordinates
$y/R$, $\psi$, $\phi$, that ${\tilde y}/R$, $\tilde \psi$
and $\tilde \phi$ are subject to the further identifications:
\be
\Bigl({{\tilde y}\over R},{\tilde \psi},{\tilde \phi}\Bigr)\sim
\Bigl({{\tilde y}\over R},{\tilde
\psi},{\tilde \phi}\Bigr)+2\pi\,l \Bigl({1\over k}, -{1\over k}, 0\Bigr)
\label{k}
\ee
with $l=0,\ldots,k-1$. The identifications above generate a group
isomorphic to $\mathbb{Z}_k$ and the space
characterized by the metric (\ref{r=0}) is topologically equivalent to
an orbifold $(\mathbb{R}^3\times S^3)/\mathbb{Z}_k$.
The orbifold action $\mathbb{Z}_k$ has fixed points where both the
$\tilde y$ and $\tilde \psi$ cycles shrink to
zero size, which happens at $r=0$ and $\theta=\pi/2$. Thus the spectral
flow metrics with $k>1$ have $\mathbb{Z}_k$
orbifold singularities of the same kind as the original 2-charge
metrics (\ref{mm}).

\subsubsection{Metrics obtained after $ST_yT_{z_1}S$ duality:  orbifold 
singularities}

Let us now turn to the case of the metrics obtained from the spectral
flow metrics by S,T dualities. Since these dualities interchange $n_1$
and $n_p$ and do not change the angular momenta $J_\psi$ and $J_\phi$,
we
find, using (\ref{gamma1}-\ref{gamma2})
\be
\gamma'_1 = {R'\over \sqrt{Q'_1 Q'_5}}\,\tg'_1= {J_\psi\over n_p
n_5}=-{k\over n_5(kn+1)}\,,\quad
\gamma'_2 = {R'\over \sqrt{Q'_1 Q'_5}}\,\tg'_2= {J_\phi\over n_p
n_5}={1\over n_5 n}
\label{dualconical}
\ee
We thus see that the parameter $\gamma'$ after duality
\be
\gamma'=\gamma'_1+\gamma'_2={1\over n_5 n (kn+1)}
\label{dualg}
\ee
is again of the form (\ref{sf}) with the integer $k$ replaced by the
integer
\be
k'=n_5 n (kn +1)\,.
\ee
In this case, then, the $\tilde y$, $\tilde \psi$ and $\tilde \phi$
coordinates can be written as
\be
{\tilde y}={y\over k'}\,,\quad {\tilde \psi}=
\psi - {1\over n_5\,n}\,{y\over R}\,,\quad
{\tilde \phi}=\phi + {k\over n_5\,(k n+1)}\,{y\over R}
\ee
and the space (\ref{r=0}) is topologically equivalent to
$(\mathbb{R}^3\times S^3)/\mathbb{Z}_{k'}$, where now
the $\mathbb{Z}_{k'}$ group acts as
\be
\Bigl({{\tilde y}\over R},{\tilde \psi},{\tilde \phi}\Bigr)\sim
\Bigl({{\tilde y}\over R},
{\tilde \psi},{\tilde \phi}\Bigr)+2\pi\,l \Bigl({1\over k'}, -{1\over
n_5\,n}, {k\over n_5\,(k n+1)}\Bigr)
\label{k'}
\ee
with $l=0,\ldots,k'-1$. Denote by $\omega$ the generator of this
$\mathbb{Z}_{k'}$ group. We notice that, in contrast with the orbifold 
action
(\ref{k}), $\omega$ acts non-trivially on all the three cycles
$\tilde y$, $\tilde \psi$ and ${\tilde \phi}$. However, the group
element $\omega^{n_5\,n}$ only acts by a $-2\pi$ rotation on the
cycle $\tilde \psi$. Thus at $r=0$ and $\theta=0$, where the other
two cycles $\tilde y$ and $\tilde \phi$ shrink, $\omega^{n_5\,n}$
has a fixed point and the 6-dimensional space (\ref{r=0}) has
an orbifold singularity. The order of this orbifold singularity
is $k'/ (n_5 n)= k n +1$. Similarly, if $k$ and $n_5$ have no
common factors, the group element $\omega^{n_5\,(k n+1)}$
acts trivially on $\tilde \phi$ and has a fixed point when
$\tilde y$ and $\tilde \psi$ shrink, which happens at $r=0$ and 
$\theta=\pi/2$.
Thus at $r=0$ and $\theta=\pi/2$ there is an orbifold singularity
of order  $k'/ (n_5 (k n+1))= n$. If $n_5$ and $k$ have a common factor
$m$, i.e. $n_5=m\,{\tilde n}_5$ and $k=m\,{\tilde k}$, then
$\omega^{{\tilde n}_5\,(k n+1)}$ has a fixed point at $r=0$ and 
$\theta=\pi/2$
and the order of the orbifold singularity increases to
  $k'/ ({\tilde n}_5 (k n+1)) = m\, n$.

In conclusion, the metrics obtained from the spectral flow
metrics by $S$ and $T$ dualities have orbifold singularities
at both $(r=0, \theta=0)$, and at $(r=0, \theta=\pi/2)$.
The order of the orbifold singularity is $k n+1$ at $(r=0,\theta=0)$
and $m\,n$ at $(r=0,\theta=\pi/2)$, where  $m$ is the highest common 
factor shared by $n_5$ and $k$.

Note: In the classical limit of the D1-D5 system we take $n_1, 
n_5\rightarrow \infty$, though BPS states exist of course for all $n_1, 
n_5$. If we start with a large $n_5$ then the orbifold shifts involving 
${1\over n_5}$ in (\ref{k'}) are very close together, and
cannot be seen in the classical limit $n_5\rightarrow\infty$.  But 
since microstates exist for all $n_1, n_5$ the geometries studied in 
this section can be considered for $n_5$ of order unity, and then they 
give well defined classical metrics with
the orbifold group (\ref{k'}).

\subsubsection{Absence of horizons and closed timelike curves}

If we write down a generic 3-charge solution with rotation, we find
closed timelike curves
(see for example \cite{her}). But we expect that geometries that
actually arise as duals to 3-charge states
will be free of pathologies. In \cite{gms} computations were developed
to show that the geometries constructed there
had no horizons and no closed timelike curves. A similar result holds
for the geometries found in this paper; the computations are given in
Appendix \ref{app2}.

\section{Wave equation for  a scalar}\setcounter{equation}{0}

In \cite{lm4},\cite{park} the wave equation for a massless minimally
coupled scalar
was studied in the 2-charge geometry.
Such a scalar arises for instance from fluctuations of the metric on
$T^4$, for example the component $h_{z_1z_2}$.
It was found that the wavepacket spent a time $\Delta t_{sugra}$
traveling down and back up the `throat' of the supergravity solution.
The time $\Delta t_{sugra}$ exactly equalled the time taken for
excitations to travel around the `effective string' in the dual CFT.
We now compute $\Delta t_{sugra}$ for the 3-charge solutions we have
found, and then use the result to find the dual state in the CFT.

The wave equation for a massless minimally coupled scalar in the 6-d
geometry is\footnote{The 6-d string metric is obtained by ignoring the
$T^4$, and the 6-d Einstein metric turns out to be the same as the 6-d
string metric.}
\be
\Box\,\Phi \equiv  {1\over \sqrt{-g}}\,\partial_\mu\,
\Bigl( \sqrt{-g}\,g^{\mu\nu}\,\partial_\nu\,\Phi\Bigr)=0
\ee
We give $g^{\mu\nu} , \det g$ in  Appendix \ref{app3}.
Writing
\be
\Phi(t,y,r,\theta, \psi,\phi,)=
\exp\Bigl(-i{\omega}\,{t\over R}+i{\lambda}\,{y\over R}+i m_1\,\psi + i
m_2\,\phi\Bigr)\,
{\tilde \Phi}(r,\theta)
\label{ansatz1}
\ee
we get a wave equation that is separable in $r,\theta$
\cite{cveticlarsen}. We write
\be
{\tilde \Phi}(r,\theta)=H(r)\,\Theta(\theta)
\label{ansatz2}
\ee
We introduce the dimensionless radial coordinate
\be
x=r^2\,{R^2\over Q_1 Q_5}
\ee
and the following convenient quantities
\bea
&&\delta = \sqrt{\eta}\,|\gamma_1+\gamma_2|=\sqrt{\eta}\,\gamma\,,\quad
\sigma^{2}= \Bigl[(\omega^2-\lambda^2){Q_1 Q_5\over
R^4}\Bigr]^{-1}\nonumber\\
&&\nu = \Bigl(1+\Lambda - (\omega^2-\lambda^2){Q_1 + Q_5\over
R^2}-(\omega-\lambda)^2 {Q_p\over R^2}\Bigr)^{1\over 2}\nonumber\\
&&\xi=\sqrt{\eta}\,\Bigl({\omega\over \eta}-\lambda\,
{Q_p (Q_1+Q_5)\over Q_1 Q_5}- m_1\,\gamma_1 - m_2\,\gamma_2
\Bigr)\nonumber\\
&&\zeta=\sqrt{\eta}\,\Bigl(\lambda + m_1\,\gamma_2 +
m_2\,\gamma_1
\Bigr)
\label{def}
\eea
Then the radial and angular part of the wave equation become
\be
4{d\over d\,x}\Bigl(x (x + \delta^2){d\over d\,x}\Bigr) H +
\Bigl[\sigma^{-2}\,x + 1-\nu^2 +{\xi^2 \over x+\delta^2}
-{\zeta^2\over x}\Bigr]\, H  =0
\ee

\bea
&&\!\!\!\!\!\!\!\!\!\!\!\!\!\!\!\!\!\!{1\over \sin 2\theta}{d\over
d\,\theta}\Bigl(\sin 2\theta {d\over d\,\theta}
\Bigr)\,\Theta+\Lambda\,\Theta+\Bigl[
-{m_1^2\over \cos^2\theta}-{m_2^2\over \sin^2\theta} \nonumber\\
&&+ {\tg_1+\tg_2\over \sigma^2}\,\eta\,
\bigl(\tg_1\, \sin^2\theta + \tg_2\,\cos^2\theta\bigr)\Bigr]\,\Theta = 0
\label{angular}
\eea
Reality of the metric implies that the wave equation is real; thus
the complex
conjugate of a solution gives another solution. We can thus take
\be
\xi\ge 0
\ee
The solutions with $\xi<0$ are obtained by complex conjugation.

\subsection{Solving the wave equation}

The radial wave equation can be solved  in the two
regions
\bea
&&\mathrm{outer~region:}\quad x\gg 1 \nonumber\\
&&\mathrm{inner~region:}\quad x\ll \sigma^2
\eea
If one chooses the frequency of the scattering wave to be very low
\be
\sigma^2\gg 1
\label{lowfrequency}
\ee
the inner and outer regions have a wide overlap, where the two
limiting solutions
can be matched. Due to this large overlapping region, we expect that a
reliable
solution can be found, in the low frequency limit (\ref{lowfrequency}),
without
the need to introduce a further region or to make any further
assumption on
the parameters. In particular the quantity
\be
(\omega^2-\lambda^2){Q_1 + Q_5\over R^2}+(\omega-\lambda)^2 {Q_p\over
R^2}
\ee
is {\it not} assumed to be small. For large $\omega$ it can be seen
from  (\ref{def}) that $\nu$ becomes imaginary; this corresponds to
energies where the quantum travels over the potential barrier in the
`neck' region instead of tunneling through it. Since the CFT is known
to describe the low energy dynamics of the system, we will restrict
ourselves to real $\nu$; we can choose the sign of $\nu$ to be positive.

Note also that in the low frequency limit (\ref{lowfrequency}) the
angular part
of the wave equation (\ref{angular}) reduces to the angular equation in
flat space and thus the eigenvalue
$\Lambda$ is
\be
\Lambda=l(l+2)
\ee

The technique of matching solutions across the two regions is well
known, and details of the computation are given in Appendix \ref{app4}.
Here we
outline the main steps and results. The solution
in the outer region is a linear combination of Bessel's functions
\be
H_\mathrm{out}={1\over\sqrt{x}}\bigl[C_1\,J_\nu(\sigma^{-1}\sqrt{x})+
C_2\,
J_{-\nu}(\sigma^{-1}\sqrt{x})\bigr]
\label{pone}
\ee
The coefficients $C_1$ and $C_2$ are fixed by demanding continuity with
the inner
solution. The latter is uniquely determined by the requirement of
regularity
at $x=0$ and
is given in terms of the hypergeometric function:
\be
H_\mathrm{in}= x^\alpha\,(x+\delta^2)^\beta\,
F\Bigl(p,q;1+2\alpha;-{x\over \delta^2}\Bigr)
\ee
with
\bea
\alpha &=& {|\zeta|\over 2\,\delta} = {\sqrt{\eta}\over
2\,\delta}\,|\lambda + m_1\,\tg_2 + m_2\,\tg_1|\nonumber\\
\beta&=&{\xi\over 2\,\delta} = {\sqrt{\eta}\over
2\,\delta}\,\Bigl({\omega\over \eta}-\lambda\,
{Q_p (Q_1+Q_5)\over Q_1 Q_5}- m_1\,\tg_1 - m_2\,\tg_2 \Bigr)\nonumber\\
p&=&{1\over 2}+\alpha+\beta+{\nu\over 2}\,,\qquad q={1\over
2}+\alpha+\beta-{\nu\over 2}
\eea
As a result of the matching we get
\be
{C_2\over C_1}={\Gamma(-\nu+1) \Gamma(-\nu)\over
\Gamma(\nu+1)\Gamma(\nu)}\,
{\Gamma({1\over 2}+
\alpha+\beta+{\nu\over 2}) \Gamma({1\over 2}+
\alpha-\beta+{\nu\over 2})\over \Gamma({1\over 2}+
\alpha+\beta-{\nu\over 2})
\Gamma({1\over 2}+ \alpha-\beta-{\nu\over 2})}\,
\Bigl({\delta^2\over 4 \sigma^2}\Bigr)^\nu
\label{result}
\ee

Note that since $\delta^2<1\ll \sigma^2$ we have $C_2\ll C_1$.

\subsection{Time of travel and absorption probability}

For very large $x$ we get from (\ref{pone})
\bea
H_\mathrm{out}&=&\sqrt{2\sigma\over\pi}{1\over
x^{3/4}}\Bigl[e^{i(\sigma^{-1}\,\sqrt{x}-{\pi\over 4})}
(C_1 e^{-i {\pi\nu\over2}} + C_2 e^{i {\pi\nu\over2}})\nonumber\\
&+& e^{-i(\sigma^{-1}\,\sqrt{x}-{\pi\over 4})}
(C_1 e^{i {\pi\nu\over2}} + C_2 e^{-i
{\pi\nu\over2}})\Bigr](1+O(x^{-{1\over 2}}))
\eea
The ratio between
the outgoing and the ingoing wave amplitude is (ignoring the constant
phase shift caused the by factors $-{\pi\over 4}$)
\bea
{\cal R}&=&{C_1 \,e^{-i{\pi\nu\over2}} + C_2\,
e^{i{\pi\nu\over2}}\over C_1 \,e^{i{\pi\nu\over2}} + C_2 \,
e^{-i{\pi\nu\over2}}} \nonumber\\
&=& e^{-i \pi\nu} + (1- e^{-2 i \pi\nu})\,\Bigl({C_2\over C_1}+
O\Bigl({C_2\over C_1}\Bigr)^2\Bigr)
\label{reflection}
\eea

In \cite{lmhot} a procedure was given to compute the travel time in the
throat from ${\cal R}$; we summarize the
method here. The  quantum coming in from infinity tunnels through the
`neck' region
with some probability $p<<1$ and enters the `throat'. Here it travels
freely down to the `cap'
and bounces back up. We again have the same probability $p$ that it
emerges  to infinity,
while with probability $1-p$ it turns
back for another trip in the throat. We thus get emergent waves at
times separated by a fixed interval $\Delta t_{sugra}$.

To find $p$ and $\Delta t_{sugra}$ we note that
${\cal R}$ can be written in the form
\be
{\cal R}= a + b\sum_{n=1}^\infty e^{2\pi i \,n {\omega\over R}\,\Delta
t}
\label{form}
\ee
where $a$ and $b$ are some real functions of $\omega$. Let us send in
from
infinity a  wave packet
\be
\int dk_r\,f(k_r)\,e^{-i{k_r\over R}r -i{\omega\over R} t }
\ee
where $k_r/R$ is the radial wave number
$k_r=\sqrt{\omega^2-\lambda^2}$.
After scattering from the geometry the wavepacket will be
\be
\int dk_r\,f(k_r)\,\Bigl[e^{-i{k_r\over R}r -i{\omega\over R} t} +
{\cal R}\,e^{i{k_r\over R}r -i{\omega\over R} t} \Bigr]
\ee
       From the form (\ref{form}) of $\cal R$ we see that
the wave packet will have peaks  at
\be
k_r\,r = \omega (t - 2\pi\, n\,\Delta t)\quad n=0,1,\ldots
\ee
The $n\ge 1$ peaks represent waves that have travelled $n$ times down
the throat and back, and we identify $\Delta t=\Delta t_{sugra}$.  From
(\ref{form}) we also see that the
probability to enter the throat and reemerge is
\be
P=p^2=|b|^2
\ee
\medskip

{\bf Note:}
In order to be able to distinguish between successive
peaks of the wavepacket, the separation between the peaks (seen at
infinity) should be larger than the
width of the wavepackets. This implies
\be
\Delta t\gg \Bigl({\Delta k_r\over R}\Bigr)^{-1}\sim {R\over k_r}
\label{boner}
\ee
     Our approximate solution (\ref{result}) is only valid in
the low frequency limit $k_r\ll R^2/\sqrt{Q_1 Q_5}$, so
we should have $\Delta t \gg \sqrt{Q_1 Q_5}/R$.
We will find that $\Delta t\sim R/ \delta$
so the requirement (\ref{boner}) is
\be
{R^2\over\delta}\gg \sqrt{Q_1 Q_5}
\label{smallgamma}
\ee
The above condition can be satisfied either by taking $R$ large or
$\delta$ small.

\medskip

We now list the results found in Appendix \ref{app5}.

\subsubsection{Energy threshold for absorption}

If the wave frequency $\omega$ is low enough so that
\be
\beta <\alpha + {\nu+1\over 2}
\label{noabs}
\ee
then the wave is reflected back at the neck region and the absorption
probability vanishes.

We can interpret this threshold in the following way. If the energy of
the quantum is low enough then we cannot
fit a complete wavelength in the `throat'. In the limit of large $R$
the throat is a large asymptotically AdS region.
The spectrum of a scalar in such an asymptotically AdS geometry was
computed in \cite{lmpp}.
Adapting those results to the notation of our paper, we find that
the energy levels are given by solutions to the equation
\be
\beta = \alpha + {l+2\over 2} + k\,,\quad k=0,1,2,\ldots
\ee
If we take the limit $R\rightarrow\infty$ then we get a large AdS
region, and we can compare
quantities to those computed in asymptotically AdS space. Taking this
limit we see
    that the frequencies (\ref{noabs}) which are not absorbed  are
those which lie below the $AdS_3$ excitation spectrum.

\subsubsection{Travel time in the geometry}

If $\omega$ is high enough so that
\be
\beta>  \alpha + {\nu+1\over 2}
\label{abs}
\ee
then the reflection amplitude computed in (\ref{refapp}) has a form
similar to (\ref{form}) with the sum over phase factors
\be
     \sum_{n=1}^\infty e^{2\pi i\,n (\beta-\alpha -{1+\nu\over 2})}\equiv
\sum_{n=1}^\infty e^{2\pi i \,n {\omega\over R}\,\Delta
t}
\label{osc}
\ee
The factor $\beta-\alpha$ depends linearly on $\omega$; the parameter
$\nu$, on the
other hand
has a non-linear dependence on $\omega$ that will distort the wave
packet.
Note however that
\be
\beta-\alpha -{1+\nu\over 2}={\xi - |\zeta| - \delta\,(1+\nu) \over
2\delta}=
\Bigl(\beta - \alpha - {l+2\over 2}\Bigr)\,(1+O(\delta\,\epsilon))
\ee
where
\be
\epsilon \equiv (l+1)- \sqrt{(l+1)^2 -
(\omega^2-\lambda^2)\,{Q_1+Q_5\over R^2}-
(\omega-\lambda)^2\,{Q_p\over R^2}}
\ee
We take all the $Q_i$ to be of the same order. We see from
(\ref{smallgamma})
that travel time makes good sense only if ${\delta }{Q_i\over R^2}<<1$.
Thus either
$\delta\ll 1$ or ${Q_i\over R^2}\ll 1$ (or both). In either case we
find that
$\delta\epsilon\ll 1$.
We then find
\be
\beta-\alpha -{1+\nu\over 2}\approx \beta - \alpha - {l+2\over
2}={\omega\over 2 \sqrt{\eta}\,\delta}
+(\omega~\mathrm{independent~terms})
\ee
In this limit the wave packet is not distorted and it travels up and
down the throat in the time
\be
\Delta t = {\pi R\over \sqrt{\eta}\,\delta} = {\pi\,R\over \eta\,\gamma}
\label{time}
\ee

\subsubsection{Absorption probability}

The probability for the wave to be absorbed and re-emitted in the
throat is found to be
\be
P= \left({4 \pi^2\over
\Gamma^2(\nu)\,\Gamma^2(\nu+1)}\right)^2\,\Bigl({\delta\over 2
\sigma}\Bigr)^{4\nu}\,\left({\Gamma({1\over 2}+
\alpha+\beta+{\nu\over 2}) \Gamma({1\over 2}+
\beta-\alpha+{\nu\over 2})\over \Gamma({1\over 2}+
\alpha+\beta-{\nu\over 2})
\Gamma({1\over 2}+ \beta-\alpha-{\nu\over
2})}\right)^2
\label{abs1}
\ee
The probability for just absorption or just emission is $p=\sqrt{P}$.

\subsubsection{The factor $\eta$ as a redshift}

In \cite{lm4} the travel time was computed for 2-charge D1-D5
geometries, and was found to be
\be
\Delta t_{sugra}={\pi R\over \gamma}
\ee
This is seen to differ by a factor $\eta$ from the 3-charge
(\ref{time}). We offer a simple physical
explanation for this factor.

Consider first the D1-D5 system. In the dual CFT the absorption of the
quantum is described by the creation
of a set of left and a set of right moving excitations, which travel at
the speed of light around the `component string' in a time
$\Delta t_{CFT}=\Delta t_{sugra}$. Now suppose we have a P charge as
well. This corresponds to the presence of left movers on the component
strings. But the left and right  excitations in the CFT travel around
the component  without interacting with each other, so one may think
that one again gets the same $\Delta t$ as in the 2-charge case and
thus the value of the P charge does not enter $\Delta t$.

But this cannot be right, since the D1, D5, P charges can all be
permuted by duality. Indeed the factor $\eta$ makes $\Delta t$ in
(\ref{time}) invariant under such permutations. To understand the role
of $\eta$ consider the  limit
of the metric (\ref{extremalmetric}) for small $r$ and small conical
defect $\delta$
\be
r\ll \sqrt{Q_i} \quad (i=1,5)\,,\quad \delta\ll {R^2\over \sqrt{Q_1
Q_5}}
\label{nh}
\ee
In this case we have a large AdS type region which would possess a CFT
dual description. In this limit $f\ll Q_i$ and thus one can replace
$H_i$ by $Q_i/f$,
obtaining an asymptotically $AdS_3\times S^3$ geometry.
As is clear from our computation above, the time of travel is
dominated by the time spent by the wave in this
part of the geometry. Let us look at the form of the metric
(\ref{extremalmetric}) in the
``near horizon'' limit (\ref{nh}). We get
\be
{d s^2_{\mathrm{n.h.}}\over \sqrt{Q_1 Q_5}}=-(\rho^2 +
\gamma^2)\,(\eta\,d{\tilde t})^2 +
{d\rho^2\over \rho^2 + \gamma^2}+\rho^2 d{\tilde y}^2 +
d\theta^2 + \cos^2\theta\,d{\tilde \psi}^2 + \sin^2\theta\,d{\tilde
\phi}^2
\ee
where we have made the following coordinate redefinitions
\bea
&&\rho^2 = {r^2\over \eta}\,{R^2\over Q_1 Q_5}\,,\quad {\tilde
t}={t\over R}\,,\quad {\tilde y}=
{1\over R}\,\Bigl[y-\eta\,{Q_p (Q_1+Q_5)\over Q_1 Q_5}t\Bigr]\nonumber\\
&&{\tilde \psi}= \psi - \eta\,\Bigl[\gamma_1 - \gamma_2 {Q_p
(Q_1+Q_5)\over
Q_1 Q_5}\Bigr]\,{t\over R} -
\gamma_2\,{y\over R}\nonumber\\
&&{\tilde \phi}= \phi - \eta\,\Bigl[\gamma_2 - \gamma_1 {Q_p
(Q_1+Q_5)\over
Q_1 Q_5}\Bigr]\,{t\over R} -
\gamma_1\,{y\over R}
\eea
As anticipated, the near horizon metric $d s^2_{\mathrm{n.h.}}$ is
locally $AdS_3\times S^3$ with curvature radius $(Q_1Q_5)^{1\over 4}$,
but the
time is rescaled by $\eta$ with respect to the time $t$ at
asymptotically flat infinity.
Thus the time computed in the CFT will be a factor of $\eta$ times the
time between wavepackets measured
at infinity. If we take a limit $R\rightarrow\infty$ keeping other
parameters fixed then $Q_p\rightarrow 0$, $ \eta\rightarrow 1$
and we can directly compare the CFT time to the gravity time. We will
take such a large $R$ limit in the more detailed analysis of the CFT
state below.

\section{Finding the CFT duals}
\label{cftdualsSec}\setcounter{equation}{0}

We started with CFT states (\ref{two}) and found their gravity
duals (\ref{extremalmetric}). We then made
new solutions by applying $S T_{y} T_{z_{1}} S$ duality to 
(\ref{extremalmetric}) and
obtained the solutions (\ref{pthree}). What are the CFT
states dual to the geometries (\ref{pthree})? We use two closely
related tools to identify the CFT duals: the time of travel
$\Delta t_{sugra}$ down the throat of the supergravity solution, and
the threshold of absorption into this throat. Longer travel times map
to longer components of the `effective string' in the CFT, and lower
absorption thresholds also reflect longer effective strings.

\subsection{Time of travel}

In \cite{lm4} it was found that for the 2-charge D1-D5 geometries
(\ref{mm}) a quantum falling into the throat emerges
after a time
\be
\Delta t_{sugra} ={\pi R\over  \gamma}
\label{oone}
\ee
or its integer multiples. In the dual CFT the corresponding state is
(\ref{2charge}), where twist operators have
joined together
\be
k={1\over \gamma}
\label{oonep}
\ee
copies of the CFT together to create `effective strings' of length
$2\pi R k={2\pi R\over \gamma}$. In the gravity picture a quantum can
fall down the throat of the geometry; in the dual CFT description the
energy of the quantum gets converted to a set of left and a set of
right moving vibrations on the `effective string' \cite{cm, dm}. These
vibrations
travel at the speed of light and meet halfway
around the effective string, so that the energy can leave the effective
string after a time
\be
\Delta t_{CFT}={1\over 2} {2\pi R k}=\pi R k
\label{otwo}
\ee
in exact agreement with (\ref{oone}). (If the vibrations fail to
collide and leave the string, then they encounter each other again
after a time (\ref{otwo}) etc., this corresponds to the successive
waves emerging at separations $\Delta t_{sugra}$ in the gravity
picture.)

Note that the number of effective strings created by the twists is
\be
m={n_1n_5\over k}={N\over k}
\label{othree}
\ee
Recall that each connected piece of the effective string is termed a
`component
string'.

\subsubsection{The number of component strings $m'$ for the 3-charge
states (\ref{pthree})}

The 3-charge geometries (\ref{extremalmetric}) had D5, D1, P charges
$(n_5, n_1, n_p)$. The orbifold CFT had  $N=n_1n_5$, the winding number
of each component string was $k$, and thus the number of component
strings was
\be
m={N\over k}={n_1n_5\over k}
\label{ofive}
\ee
By the duality $S T_{y} T_{z_{1}} S$ we obtained the geometries 
(\ref{pthree}) which had
charges\footnote{S,T dualities certainly map a state of the 3-charge
system to another state of the system, but after duality we may not be
in a range of parameters where the state is well described by the {\it
conformal}  field theory.  The conformal limit is the low energy limit,
and is attained for small ${\sqrt{Q_1Q_5}\gamma\over R}$. If we start
with a large circle radius $R$ then after dualities we get  small
$R'$. But since we are dealing with BPS states we can follow the state
as we increase $R'$ and get back to a CFT domain. It is this latter CFT
state that we will mean when we look at the states after $S T_{y} 
T_{z_{1}} S$ duality.}
\be
n_5'=n_5, ~~n_1'=n_p, ~~n_p'=n_1
\ee
The orbifold CFT now has
\be
N'=n_1'n_5'=n_5n_p
\ee
The travel time for such geometries is derived in (\ref{time}). Take
the limit $R'\rightarrow\infty$; this gives
a large AdS region which will be the dual of the CFT.
We get $Q'_p\rightarrow0$, $\eta'\rightarrow 1$ and the travel
time is
\be
\Delta t_{sugra}={\pi R'\over \gamma'}=\pi R' n_5 n (n k+1)
\ee
where we used (\ref{dualg}). Thus we expect the winding number of each
component string to be
\be
k'=k {n_p\over n_1}=n_5 n (n k+1)
\label{ieight}
\ee
The number of components of the effective string are then
\be
m'={N'\over k'}={n_5n_p\over k}{n_1\over n_p}={n_1n_5\over k}
\ee
We thus see that while $N, k$ change to $N', k'$, {\it the number of
components of the effective string remains unchanged}
\be
m'=m
\ee
To interpret this fact we recall the physical significance of $m$ found
in \cite{lm4}. When we put a particle in the throat of the gravity
solution then we excite left and right movers on one component string
in the dual CFT. Putting another particle in the throat excites another
component string, and so on. Let each particle have the longest
possible wavelength which can still fit in the throat of the geometry.
When we have enough particles in the throat so that we use up all the
$m$ component strings in the CFT then we find that we have enough
energy in the supergravity solution to give a backreaction of order
unity in the geometry; the geometry distorts and we can no longer study
the particles as independent excitations. We now see that under S,T
dualities this critical number of particles for the geometry
stays unchanged.

\subsubsection{Level of excitation of the component strings}

The CFT state for the geometries (\ref{extremalmetric}) had $n_p$ units
of momentum distributed equally over $m$ component strings; thus each
component had
\be
T={n_p\over m}
\ee
units of momentum. It will be helpful to write this in terms of the
basic units of momentum on the component string. Since the string has
winding number $k$ the excitations come in units of ${1\over k}$. Thus
the number of these basic units of momentum on each component string is
\be
\hat T=kT={n_p k\over m}={n_p\over m}{N\over m}={n_1n_5n_p\over m^2}
\ee
For the state after $S T_{y} T_{z_{1}} S$ duality we have $n'_p=n_1$ 
units of momentum,
distributed over $m'=m$ component strings, and
\be
T'={n_1\over m}, ~~~\hat T'=k' T={n_1 \over m}k{n_p\over
n_1}={n_1n_5n_p\over m^2}
\label{isix}
\ee
so we see that
\be
\hat T'=\hat T={n_1n_5n_p\over m^2}
\ee
Thus $\hat T$ is a duality invariant; the charges permute under
dualities and we have seen that $m$ stays unchanged.
We note that the number of possible states on each component string
depends on the number $\hat T$: We have to just excite free bosons and
fermions on the component string so that the total level (as measured
in units of the basic excitation
on the component string) is $\hat T$. Thus we see that under $S T_{y} 
T_{z_{1}} S$
duality the number of allowed states on the component strings remains
unchanged.

\subsubsection{The state after $S T_{y} T_{z_{1}} S$ duality}

The angular momentum of the state does not change under the $S T_{y} 
T_{z_{1}} S$
duality
(the T dualities are all along  compact directions, while angular
momentum reflects the properties of the state in the noncompact
directions).  For the entire state we have from (\ref{twop})
\be
j=-{n_1n_5\over 2}(2n+{1\over k}), ~~~\bar j=-{n_1n_5\over 2k}
\ee
Both before and after the $S T_{y} T_{z_{1}} S$ duality we have $m$ 
component strings,
so
the angular momentum on each component string is
\be
\hat j=-{1\over 2}- n k\,, ~~~\hat {\bar j}=-{1\over 2}
\ee
Both before and after the duality we have the same `level' of
excitation $\hat T$ on each component string. The state before the
duality was given by
(\ref{sstate})  -- the charge $(j, \bar j)$ was attained by the lowest
energy possible by having all fermion spins aligned and all fermions in
the lowest levels allowed by the Pauli exclusion principle. We see that
the only way to make the state after $S T_{y} T_{z_{1}} S$ duality is 
to make a
construction similar to (\ref{sstate})
\be
(\prod [J^-_{-2n}J^-_{-2n{k'-1\over k'}}\dots J^-_{-{2\over
k'}}])|\Psi^{--}(k',0)\rangle
\label{sstatep}
\ee
The component strings now have winding $k'$ each instead of $k$, but
the rest of the construction is the same.

\subsection{Absorption threshold}

In section 5 it was found that an incoming wave was absorbed only if
\be
\beta-\alpha>{\nu+1\over 2}
\label{ione}
\ee
Take the CFT limit $R\rightarrow\infty$ which gives
$Q_p\rightarrow 0, \eta\rightarrow 1$. The angular momentum components
$m_1, m_2$ imply for the two $su(2)$ factors the eigenvalues
\be
m={m_1-m_2\over 2}, ~~~\bar m=-{m_1+m_2\over 2}
\ee
The condition (\ref{ione}) then gives a pair of conditions
from the two possible signs of the absolute value in $\alpha$
\bea
\omega-2(m {j\over N}+\bar m{\bar j\over N})-\lambda+2(m{j\over N}-\bar
m{\bar j\over N})&>&\gamma(l+2)\nonumber \\
\omega-2(m {j\over N}+\bar m{\bar j\over N})+\lambda-2(m{j\over N}-\bar
m{\bar j\over N})&>&\gamma(l+2)
\label{itwo}
\eea
Here $(j, \bar j)$ are the angular momenta of the geometry into which
the quantum is falling
\be
{j\over N}=-(n+{\gamma\over 2}), ~~~{\bar j\over N}=-{\gamma\over 2}
\ee
Note that ${\omega\pm\lambda\over 2}$ are the increments of $L_0, \bar
L_0$ caused by the incoming quantum, so we can write (\ref{itwo}) as
\bea
\Delta h&>&\gamma+({l\over 2}-m)\gamma-2mn \label{ithree}\\
\Delta \bar h&>&\gamma+({l\over 2}-{\bar m})\gamma
\label{ifour}
\eea

\subsubsection{Absorption of quanta in the CFT description}

In \cite{cm, dm, angular} the CFT description of absorption was
studied. We have supposed that the quantum being absorbed
is a scalar arising from the component $h_{ij}$ of the metric on the
$T^4$. The quantum has angular momentum
$l$, which means that it is in the representation $({l\over 2}, {l\over
2})$ of $su(2)_L\times su(2)_R$. In the CFT state this
quantum creates an excitation that has
\be
\partial X, \psi \dots \psi, ~~~~~\bar \partial X, \bar\psi, \dots
\bar\psi
\ee
where the $X$ variables carry the indices $i,j$ (we must symmetrize
over the two permutations) and there are $l$ fermions on each of the
left and right sectors. The fermions are in the Ramond sector, and thus
both fermionic and bosonic excitations come in multiples of the basic
harmonic on the component strings
\be
\Delta h=\Delta \bar h={1\over k}=\gamma
\ee
The $(J^3, \bar J^3)$ quantum numbers in $su(2)_L\times su(2)_R$ are
$(m, \bar m)$. There are two species of left moving fermions with
$m={1\over 2}$ and two with $m=-{1\over 2}$; similarly for the right
movers. We will simply write $\psi^\pm, \bar \psi^\pm$ without
distinguishing the two species since the difference is irrelevant for
the discussion below. Thus on the left sector we must have
${l\over 2}+m$ fermions $\psi^+$ and ${l\over 2}-m$ fermions $\psi^-$.
On the right sector we
have ${l\over 2}+\bar m$ fermions $\bar\psi^+$ and ${l\over 2}-\bar m$
fermions $\bar \psi^-$.

\subsubsection{Identifying the excitations}

Recall the structure of the CFT state (\ref{sstate}) into which we are
absorbing the quantum. On each component string we have fermion zero
modes. For the left movers we started by choosing the state which is
killed by the $\psi^-_0$, and then applied operators $J^-$  which
resulted in the application of modes   $\psi^-_{-{1\over k}}\dots
\psi^-_{-n}$ (for both species of $\psi^-$). For the right movers we
also have the vacuum killed by the ${\bar\psi}^-_0$, but applied no
other
excitations.

Consider first the right movers.  The excitation $\bar\partial X$ needs
a minimum $\Delta \bar h$ of ${1\over k}=\gamma$; this is the first
term on the RHS of (\ref{ifour}). Now consider the fermions. Suppose
that $\bar m={l\over 2}$. Then we have $l$ operators ${\bar\psi}^+$
acting on
the state of the CFT. But we can choose each of these operators to be a
zero mode $\bar\psi^+_0$ which changes the vacuum on a component string
to one that is killed by $\bar \psi^+_0$. The fermions then do not
contribute to $\Delta \bar h$ and we find exact agreement with
(\ref{ifour}). Note that we could find at most two such zero modes on
any given component string, so we will have to apply the $\bar\psi^+_0$
in general to many different component strings.\footnote{We can read
off the value of $\gamma={1\over k}$ from the classical geometry, for
example by the conical defect angles of the metrics
(\ref{extremalmetric}). But $n_1, n_5$ are infinite in the classical
limit, and so $m={n_1n_5\over k}$ is also infinite in this limit. Thus
the absorption thresholds computed from the classical geometry have  an
essentially infinite number of component strings, and we will not `run
out of component strings' for any value of $l$ that we choose.}

Now suppose that ${\bar m}={l\over 2}-1$. We have $l-1$ operators $\bar
\psi^+$ and one $\bar\psi^-$. The $\bar\psi^+$ again give no
contribution to $\Delta \bar h$, but the lowest allowed mode for
$\bar\psi^-$ is $\bar\psi^-_{-{1\over k}}$ and we again get agreement
with (\ref{ifour}). Proceeding this way, we find agreement for all
$\bar m$ for the right movers.

Now consider the left movers, and let $m={l\over 2}$. We have to apply
$l$ operators $\psi^+$. The lowest excitation results if we use the
operators $\psi^+$ to {\it annihilate} $l$ modes $\psi^-_{-n}$, which
gives a total $\Delta h={1\over k}-nl$ (including the boson $\partial
X$) which agrees with (\ref{ithree}). Next consider $m={l\over 2}-1$.
Now we have $l-1$ operators $\psi^+$ which again annihilate modes
$\psi^-_{-n}$ and one operator $\psi^-$ which creates the lowest
allowed mode $\psi^-_{-(n+{1\over k})}$, again in exact agreement with
(\ref{ithree}). (The operator $\psi^-$ cannot just fill in one of the
empty levels created by annihilation of modes $\psi^-_{-n}$ since the
overall operator describing the absorption is taken to be normal
ordered.)

We thus see that the absorption threshold seen in the wave equation
analysis can be understood in detail in terms of the occupied levels on
the effective strings. This computation  supports the conjecture that
the state after duality
has the form (\ref{sstatep}) analogous to the states before the $S 
T_{y} T_{z_{1}} S$
duality; the absorption computation applies equally to both sets of
geometries.

\subsection{A puzzle about the orbifold theory}

Consider the CFT state (\ref{sstatep}) dual to the geometries obtained
after $S T_{y} T_{z_{1}} S$ duality.  The momentum $T'$ per component 
string (given in
(\ref{isix})) is {\it fractional} in general, not an integer.
Interestingly, when this momentum is measured in units of the basic
excitation ${1\over k}$ on the component string then we get an {\it
integer} $\hat T'$.
(If $\hat T'$ were nonintegral, we would have a severe contradiction,
since we could not carry the excitation on the component string.)

But here we face a puzzle, since in the orbifold CFT the quantity $T'$
is {\it also} required to be integral. The reason for this is as
follows. We have to orbifold by the symmetric group $S_{N'}$ which
permutes the $N'$ copies of the $c=6$ CFT.
In a given state of the CFT we can label the copies by how they make up
the different component strings, and even inside each component string
the copies can be ordered by the sequence in which they link up to make
the `long cycle'. But one part of the symmetry group still survives: we
can cyclically permute the copies inside a component string
\be
c_1\rightarrow c_2\rightarrow \dots \rightarrow c_k\rightarrow c_1
\label{iseven}
\ee
This symmetry forces the momenta on the component string to be integral.

Faced with this problem, we first review the steps that led us to our
solutions. The fundamental string (F)
is an elementary excitation of the theory on $M_{4,1}\times S^1\times
T^4$, so the 2-charge FP solutions
we started with certainly correspond to BPS states of the full string
theory. Since S,T dualities are exact symmetries
of the theory the 2-charge D1-D5 states are also states of the theory.
Spectral flow was just a coordinate change,
and so the 3-charge solutions that were obtained by spectral flow must
also be valid. Finally, the exact symmetry $S T_{y} T_{z_{1}} S$
was used to get the solutions which we are now discussing, so we
conclude that they must be allowed states
of the string theory.

Even though one may accept the gravity solutions one may question the
identification of the CFT state. We have made each component string
have the same winding number $k'$, but if we allowed the winding
numbers to be different then we could have carried the momentum
$n'_p=n_1$ on the component strings in such a way that each component
string had an integral number $T'$ of units of momentum. But let us
recall the evidence we have that the component strings should all be
equal:
(a) We have learnt from the explicit construction of 2-charge systems
that states with axial symmetry have all component strings identical;
if the component strings are different then all symmetries are broken
in general (b) The return time from the throat gave a precise value
that could be put in correspondence with the length of the component
strings; if component strings had different lengths then we would get
distortion of the wavepacket since different parts would be returned
after different times (c) The threshold of absorption worked out
exactly -- if we change the structure of the filled levels on the
component strings then the allowed levels that could be excited by the
incoming quantum would change. For all these reasons it seems hard to
have any other construction of the CFT state.

It may be that we need changes in our understanding of the CFT dual to
the gravity theory.
It is not completely clear where the orbifold point sits in the moduli
space, though there are some leading candidates
\cite{larsenmartinec, strominger}. It is also not clear if the orbifold
$(T^4)^N/S_N$ does describe some point in the moduli space, or if we
need to consider other related orbifold theories like the iterated
orbifolds \cite{moore}.

Before concluding we note a point about the winding number of the
component strings. The geometries
(\ref{extremalmetric}) had an integral $k$, since this number can be
traced back to the number of turns of the helix of the F string in the
starting FP solution that led to (\ref{extremalmetric}). For the
geometries obtained after $S T_{y} T_{z_{1}} S$ duality
we still found in eq.(\ref{ieight}) that the winding number of each
component string $k'$ was integral. This was important, since if $k'$
turned out fractional we could make no sense of the CFT state. But if
we assume that the quantities $T$ do not need to be integral and only
the $\hat T$ need to be integral in general, then after $S T_{y} 
T_{z_{1}} S$ duality we
would get fractional $k'$ in general. What then are the rules for the
allowed CFT states?

For the D1-D5 CFT we can have   $m$ component strings with equal
winding if
\be
{n_1n_5\over m}\in \mathbb{Z}
\ee
Given that $m$ was found to be a duality invariant, and that $n_1, n_5,
n_p$ permute under duality, we conjecture that the state can have $m$
equal length component strings if all the following conditions are true
\be
{n_1n_5\over m}\in \mathbb{Z},~~{n_5n_p\over m}\in
\mathbb{Z},~~{n_pn_1\over m}\in \mathbb{Z},~~\hat T={n_1n_5n_p\over
m^2}\in\mathbb{Z}
\ee
where we have included the requirement that $\hat T$ be integral. We
hope to return to these issues elsewhere.

\section{Discussion}\setcounter{equation}{0}

Our basic conjecture states that the black hole interior is not `empty
space with a central singularity' but a `fuzzball' of horizon size. If
we consider extremal holes, and   look at states where in the dual CFT
we have many component strings in the same state, then we can have a
good description of the geometry in classical supergravity. All
2-charge states could be approached through such classical geometries,
and in \cite{mss, gms} some classes of 3-charge extremal states were
considered and their dual geometries identified; the geometries were
smoothly capped as in Fig.1(b). In the present paper we have looked at
two families of states and their dual geometries. The geometries were
again found  to be `capped'; there were no horizons or closed timelike 
curves.

The first family arose from spectral flow of a subfamily of 2-charge
states. These 2-charge states generically had an orbifold singularity 
along a
curve; this was understood as a `trivial' singularity in the sense that
it arose from the coincidence of two or more `KK monopoles tubes' and
was thus arose only as a limit in a family of regular solutions. Since
spectral flow is given by a coordinate transformation in $AdS_3\times
S^3$ \cite{bal,mm} it is not surprising that a similar orbifold 
singularity
arose also for the 3-charge states obtained by spectral flow of the
2-charge states. What was interesting to note was that the conical
defect angle did not get corrected when the asymptotically AdS solution
is modified to become an asymptotically flat solution. If the conical
defect parameter had changed away from the form ${1\over k}$ to say an
irrational value then we would not be able to understand the orbifold
singularity in any simple way as a limit of nonsingular
solutions.

The second family we considered was found by applying S,T dualities to
the first family of  geometries so that the D1 and P charges got
interchanged.
We identified the CFT states dual to this second family of geometries
by computing the time of travel $\Delta t_{sugra}$
for a quantum to fall down the throat and bounce back out. The winding
number of the `component strings' in the CFT, computed by this method,
gave a minimum  threshold energy for  excitations of the CFT state.
This minimum energy agreed with the threshold of energy below which the
incident quantum was unable to enter the throat of the geometry, thus
confirming the identification of the CFT state. We noted that the CFT
states found this way had a fractional momentum on each `component
string' (the total momentum was of course an integer). This suggested
that we need to go beyond the simple orbifold CFT to understand all
3-charge bound states; we may need to understand deformations away from
the orbifold point  \cite{dijk} or perhaps we may have to consider
iterated orbifolds \cite{moore}.

The conical defect in the 2-charge D1-D5 geometries could be directly 
linked to the
winding number $k$ of each component string. For the metrics obtained 
by spectral flow
from 2-charge geometries we have  a similar relation since spectral 
flow is just a coordinate transformation on the geometry \cite{bal,mm}. 
The conical defect is caused by an orbifold
singularity of order $k$ along the $S^1$ given by $(r=0, \theta=\pi/2)$ 
and the component strings in the dual CFT state  have winding number 
$k$. For the geometries obtained after
S,T dualities the situation is more complex -- we have orbifold 
singularities along two different $S^1$ curves, with order
$kn+1$ at
$(r=0, \theta=0)$ and order $mn$ at $(r=0, \theta=\pi/2)$ ($m$ is the 
highest common factor shared by $n_5, k$). There is a suggestive 
pattern though that we observe. It was argued in \cite{dijk} that the 
D1-D5 CFT with charges $(n_1, n_5)$ could be mapped  to the theory with 
charges $(n_1n_5, 1)$, and that the orbifold point occurs at this 
latter value of the charges. If we set $n_5=1$  to enable a comparison 
to the orbifold theory, then we observe that the product of the orders 
of the two orbifold groups is $(kn+1)n$ which in this case equals  $k'$ 
, the winding number of the effective string in the dual CFT state.

We have noted that the travel time is symmetric between the three
charges of the solution, but the factor $\eta$ enforcing this symmetry
is responsible for providing a `redshift' which relates the time at
infinity to the time in the AdS region.
The AdS region is the one that is dual to the CFT description, so
understanding such effects in more detail may help us to
identify the deformations in the CFT that correspond to deforming AdS
space to flat space at infinity.

\section*{Acknowledgements}

S.G. was supported by  an I.N.F.N. fellowship. The work of S.D.M  was
supported in part by DOE grant DE-FG02-91ER-40690. We
thank  Yogesh Srivastava for helpful discussions.

\appendix
\section{Non-extremal rotating 3-charge metrics and their extremal
limits}\label{app1}

\renewcommand{\theequation}{A.\arabic{equation}}
\setcounter{equation}{0}

The metric for general rotating 3-charge solutions was given in
\cite{cy}, and the metric, 2-form field and dilaton
were given in \cite{gms} by generating the solution by a different
technique. We have
\bea
ds^2 &=& - \left( 1- \frac{M\cosh^2\delta_{p}}{f}
\right)\frac{dt^2}{\sqrt{H_{1}H_{5}}}  + \left(1+ \frac{M
\sinh^2\delta_{p}}{f}\right) \frac{ dy^2}{\sqrt{H_{1}H_{5}}}\nonumber\\
&-&\frac{M
\sinh2\delta_{p}}{f\sqrt{H_{1}H_{5}}} dt dy
+    f\sqrt{H_{1}H_{5}}\left(\frac{ r^2  dr^2
}{(r^2+a_{1}^2)(r^2+a_{2}^2)-M r^2} +  d\theta^2\right) \nonumber \\
          &+& \left[ (r^2+a_{1}^2) \sqrt{H_{1}H_{5}} +
\frac{(a_{2}^2-a_{1}^2)
K_{1}K_{5} \cos^2\theta}{\sqrt{H_{1}H_{5}}} \right] \cos^2\theta
d\psi^2 \nonumber \\
          &+& \left[ (r^2+a_{2}^2) \sqrt{H_{1}H_{5}} +
\frac{(a_{1}^2-a_{2}^2)  K_{1}K_{5} \sin^2\theta}{\sqrt{H_{1}H_{5}}}
\right] \sin^2\theta d\phi^2 \nonumber \\
&+& \frac{M  }{f\sqrt{H_{1}H_{5}}} (a_{1}\cos^2\theta
d\psi+a_{2}\sin^2\theta d\phi)^{2} \nonumber \\
&+& \left.\frac{2 M \cos^2\theta}{f\sqrt{H_{1}H_{5}} }\right[
\left(a_{1}\cosh\delta_{1}\cosh\delta_{5}\cosh\delta_{p} - a_{2}
\sinh\delta_{1}\sinh\delta_{5}\sinh\delta_{p}\right)  dt   \nonumber
\\
& & +  (a_{2}\sinh\delta_{1}\sinh\delta_{5}\cosh\delta_{p}-
a_{1}\cosh\delta_{1}\cosh\delta_{5}\sinh\delta_{p})dy\mbox{\huge
]}d\psi \nonumber \\
&+& \left. \frac{2 M \sin^2\theta}{f\sqrt{H_{1}H_{5}} }\right[
\left(a_{2}\cosh\delta_{1}\cosh\delta_{5}\cosh\delta_{p} - a_{1}
\sinh\delta_{1}\sinh\delta_{5}\sinh\delta_{p}\right)  dt   \nonumber
\\
& &  + (a_{1}\sinh\delta_{1}\sinh\delta_{5}\cosh\delta_{p}-
a_{2}\cosh\delta_{1}\cosh\delta_{5}\sinh\delta_{p})dy\mbox{\huge
]}d\phi \nonumber \\
&+&\sqrt{H_1\over H_5}\sum_{i=1}^4 dx_i^2\nonumber\\
C_{2} &=& \left.\frac{M \cos^2\theta}{fH_{1}} \right[ (a_{2}
\cosh\delta_{1}\sinh\delta_{5}\cosh\delta_{p} -
a_{1}\sinh\delta_{1}\cosh\delta_{5}\sinh\delta_{p}) dt \nonumber \\
& & +( a_{1}\sinh\delta_{1}\cosh\delta_{5}\cosh\delta_{p} -
a_{2}\cosh\delta_{1}\sinh\delta_{5}\sinh\delta_{p})dy \mbox{\huge
]}\wedge d\psi \nonumber \\
&+& \left.\frac{M\sin^2\theta }{f H_{1}} \right[ (a_{1}
\cosh\delta_{1}\sinh\delta_{5}\cosh\delta_{p} -
a_{2}\sinh\delta_{1}\cosh\delta_{5}\sinh\delta_{p}) dt \nonumber \\
& & +( a_{2}\sinh\delta_{1}\cosh\delta_{5}\cosh\delta_{p} -
a_{1}\cosh\delta_{1}\sinh\delta_{5}\sinh\delta_{p})dy \mbox{\huge
]}\wedge  d\phi \nonumber \\
          &-& \frac{M \sinh2\delta_{1} }{2fH_{1}}  dt\wedge dy - \frac{M
\sinh
2\delta_{5}}{2f
H_{1}}\left(r^2+a_{2}^2+M\sinh^2\delta_{1}\right)\cos^2\theta
d\psi\wedge d\phi \nonumber \\
e^{2\Phi} &=& \frac{H_{1}}{H_{5} } \label{ne}
\eea
Here
\bea
&&f=r^2+a_1^2\sin^2\theta+a_2^2\cos^2\theta\nonumber\\
&&H_i\equiv 1+K_i = 1+{M\sinh^2\delta_i\over f}, ~~~i=1,5
\eea

In terms of the parameters appearing in the metric, the D1, D5 and
momentum charges and the two angular momenta are
\bea
&&\!\!\!\!\!\!\!\!\!\!\!\!\!Q_1={M\over 2}\sinh 2\delta_1\,,\quad
Q_5 = {M\over 2}\sinh 2\delta_p\,,\quad
Q_p = {M\over 2}\sinh 2\delta_p\\
&&\!\!\!\!\!\!\!\!\!\!\!\!\!J_\psi = -M \bigl(a_{1} \cosh\delta_{1}
\cosh\delta_{5} \cosh\delta_{p} -
a_{2} \sinh\delta_{1}\sinh\delta_{5}\sinh\delta_{p} \bigr){\pi\over 4
G^{(5)}}\nonumber\\
&&\qquad= \tg_1\,{\pi\sqrt{Q_1 Q_5}\over 4 G^{(5)}}\nonumber\\
&&\!\!\!\!\!\!\!\!\!\!\!\!\!J_\phi= - M\bigl(a_{2} \cosh\delta_{1}
\cosh\delta_{5} \cosh\delta_{p} -
a_{1} \sinh\delta_{1}\sinh\delta_{5}\sinh\delta_{p} \bigr){\pi\over 4
G^{(5)}}
\nonumber\\
&&\qquad=\tg_2\,{\pi\sqrt{Q_1 Q_5}\over 4 G^{(5)}}
\eea
with $G^{(5)}$ the 5-D Newton's constant. The extremal limit is the
limit in
which $M\to 0$ while $Q_1$, $Q_5$, $Q_p$, $\tg_1$, $\tg_2$ are
kept
finite. We will give the extremal metric for generic values of the
charges and
of the angular momenta satisfying
\be
(\tg_1-\tg_2)^2-4\,Q_p\ge 0
\ee
In this case we find it convenient to parametrize $Q_p$ as
\be
Q_p=\Bigl( {\tg_1-\tg_2\over
2}\Bigr)^2-\Bigl({\tg_1+\tg_2\over 2
\mu}\Bigr)^2\,,\quad \mu>0
\label{btwo}
\ee
The extremal metric, Ramond field and dilaton are
\bea
ds^2 & = & -\frac{1}{h} (dt^2-dy^2) + \frac{Q_{p}}{h
f}\left(dt-dy\right)^{2}+ hf \left( \frac{dr^2}{r^2 +
\mu^{-1}\,(\tg_1+\tg_2)^2\,\eta} +
d\theta^2 \right) \nonumber \\
        &+& h \Bigl( r^2 + (\tg_1+\tg_2)\,\eta\,
{(1+\mu)\tg_1+(1-\mu)\tg_2\over 2\mu}-
\frac{(\tg_1^2-\tg_2^2)\,\eta\,Q_{1}Q_{5} \cos^2\theta}{h^2 f^2}
\Bigr) \cos^2\theta d\psi^2  \nonumber \\
&+& h\Bigl( r^2 + (\tg_1+\tg_2)\,\eta\,
{(1-\mu)\tg_1+(1+\mu)\tg_2\over 2\mu}\nonumber +
\frac{(\tg_1^2-\tg_2^2)\,\eta\,Q_{1}Q_{5} \sin^2\theta}{h^{2}
f^{2} }
\Bigr) \sin^2\theta d\phi^2  \nonumber \\
&+& \frac{Q_p\,(\tg_1+\tg_2)^2\,\eta^2}{h f}
\left( \cos^2\theta d\psi + \sin^2\theta d\phi \right)^{2}  \nonumber\\
&-& \frac{2 \sqrt{Q_{1}Q_{5}} }{hf}
\left[\tg_1 \cos^2\theta d\psi + \tg_2 \sin^2\theta d\phi\right]
(dt-dy)
\nonumber \\
&-& \frac{2 (\tg_1+\tg_2)\,\eta \sqrt{Q_{1}Q_{5}}}{h f}
\left[ \cos^2\theta d\psi + \sin^2\theta d\phi \right] dy +
\sqrt{\frac{H_{1}}{H_{5}}} \sum_{i=1}^{4} dx_{i}^2
\label{genericem}
\eea
\bea
C_{2} &=& - \frac{\sqrt{Q_{1} Q_{5}} \cos^2\theta}{H_1 f }
\left(\tg_2\,dt + \tg_1\,dy \right)
\wedge d\psi - \frac{\sqrt{Q_{1}Q_{5} } \sin^2\theta }{H_1 f }
\left(\tg_1\,dt + \tg_2\,dy \right) \wedge d\phi\\
&+& \frac{(\tg_1+\tg_2)\,\eta\,Q_{p} }{\sqrt{Q_{1}Q_{5}}H_1 f }
\left( Q_{1}\,dt + Q_{5}\,dy \right)
\wedge \left( \cos^2\theta d\psi  + \sin^2\theta d\phi \right)
\nonumber \\
&-& \frac{ Q_{1} }{H_1 f} dt \wedge dy - \frac{ Q_{5}\cos^2\theta
}{H_1 f} \left( r^{2} +
(\tg_1+\tg_2)\,\eta\,{(1-\mu)\tg_1+(1+\mu)\tg_2\over 2\mu}
+ Q_{1} \right)  d\psi \wedge d\phi\nonumber
\eea
\be
e^{2\Phi} = \frac{H_{1}}{H_{5}}
\ee
\bea
&&\!\!\!\!\!\!\!\!\!\!\!\!f = r^2+(\tg_1+\tg_2)\eta
\Bigl[{(1+\mu)\tg_1+(1-\mu)\tg_2\over 2\mu}\,\sin^2\theta +
{(1-\mu)\tg_1+(1+\mu)\tg_2\over 2\mu}\,\cos^2\theta\Bigr]
\nonumber\\
&&\!\!\!\!\!\!\!\!\!\!\!\! H_{1} = 1+ \frac{Q_{1}}{f}\,,\quad
H_{5} = 1+ \frac{Q_{5}}{f}\,,\quad h = \sqrt{H_{1} H_{5}}
\eea
The metric (\ref{genericem}) with $\mu\not=1$ can be obtained from
the metric (\ref{extremalmetric}), which corresponds to the case
$\mu=1$,
via a boost in the y direction:
\be
t\to t\cosh\,\delta + y\sinh\,\delta\,,\quad y\to y\cosh\,\delta +
t\sinh\,\delta
\ee
with
\be
e^{2\delta}={\mu-1\over \eta}+1
\ee

\section{Singularities, closed timelike curves and horizons}\label{app2}
\renewcommand{\theequation}{B.\arabic{equation}}
\setcounter{equation}{0}

The determinant of the metric (\ref{genericem})
\be
\sqrt{-g} = h f\,\sin\,\theta\,\cos\,\theta\,r
\ee
only vanishes at $r=0$ and $\theta=0,\pi/2$.  Note that
$\theta=0,\pi/2$ are the
points where spherical coordinates degenerate, so singularities
of (\ref{genericem}) can only occur at $r=0$. In a neighborhood of $r=0$
$y$ can be decoupled from all the other coordinates by the coordinate
change
\bea
&&{\tilde t}= t - c_t\,y\nonumber\\
&& {\tilde \psi} = \psi - c_\psi\,y\nonumber\\
&&{\tilde \phi}= \phi- c_\phi \,y
\label{genericct0}
\eea
with
\bea
&&\!\!\!\!\!\!\!\!\!\!\!\!c_t=-{(\mu-1)[\mu^2
(\tg_1-\tg_2)^2-(\tg_1+\tg_2)^2+2\mu^2 \sqrt{Q_1 Q_5}]
\over (\mu-1)
[\tg_1 (1+\mu) + \tg_2 (1-\mu)][\tg_1 (1-\mu) + \tg_2
(1+\mu)] +2 \mu^2 (1+\mu) \sqrt{Q_1 Q_5}}
\nonumber\\
&&\!\!\!\!\!\!\!\!\!\!\!\!c_\psi= 2 {\mu^2 [\tg_1(1-\mu)+\tg_2
(1+\mu)] \over (\mu-1)
[\tg_1 (1+\mu) + \tg_2 (1-\mu)][\tg_1 (1-\mu) + \tg_2
(1+\mu)] +2 \mu^2 (1+\mu) \sqrt{Q_1 Q_5}}\nonumber\\
&&\!\!\!\!\!\!\!\!\!\!\!\!c_\phi=  2 {\mu^2 [\tg_1(1+\mu)+\tg_2
(1-\mu)] \over (\mu-1)
[\tg_1 (1+\mu) + \tg_2 (1-\mu)][\tg_1 (1-\mu) + \tg_2
(1+\mu)] +2 \mu^2 (1+\mu) \sqrt{Q_1 Q_5}}
\nonumber
\label{genericct}
\eea
In the new coordinates, the part of the metric that involves $y$ and
$r$ is
\be
ds_{r-y}^2 =  {hf\over (\tg_1+\tg_2)^2\,\eta}\,\Bigl(dr^2 +
c_y^2\, r^2\,dy^2\Bigr)+ O(r^4)
\label{genericry}
\ee
with
\bea
&&\!\!\!\!\!\!\!\!\!c_y={4(\tg_1+\tg_2)\mu^2\over (\mu-1)
[\tg_1 (1+\mu) + \tg_2 (1-\mu)][\tg_1 (1-\mu) + \tg_2
(1+\mu)] +2 \mu^2 (1+\mu)
\sqrt{Q_1 Q_5}}\nonumber\\
\eea
     {}From the expressions above we see that, unless $\mu=1$, the
transformation (\ref{genericct}) involves a shift of $t$ by the periodic
coordinate $y$.
At $r=0$ the $y$ circle shrinks. Thus, as explained in
\cite{gms}, metrics with $\mu\not=1$ have closed timelike curves.

The geometries (\ref{extremalmetric}) dual to CFT states have $\mu=1$.
(The condition $\mu=1$ is seen to give, using
(\ref{btwo}), the relation $Q_p=\tilde\gamma_1\tilde\gamma_2$.)
For these metrics one can show that there are no closed timelike curves
by looking at
the determinant ${\tilde g}$ of the metric restricted to the three
periodic coordinates $y$, $\psi$ and $\phi$:
\be
{\tilde g} = {r^2\sin^2\theta\cos^2\theta\over \sqrt{(Q_1+f)(Q_5+f)}}\,
\Bigl[(r^2+ (\tg_1+\tg_2)^2\,\eta)
(f+Q_1+Q_5+Q_p)+{Q_1 Q_5\over \eta}\Bigr]
\ee
Using an argument given in \cite{gms}, it is enough to show that
the determinant above never vanishes to prove that the metric is
free of closed timelike curves. By explicit inspection, we know
that the zeros of ${\tilde g}$ at $r=0$ or $\theta=0,\pi/2$
do not signal the presence of closed timelike curves. So we need to
show that ${\tilde g}$ has no other zeros. This follows from the fact
that $f+Q_i>0$ for $i=1,5,p$. In order to prove this last statement,
consider
first the geometries obtained by spectral flow, which have
\be
\tg_1=-a\,n\,,\quad \tg_2=a\,(n+\gamma)\,,\quad Q_p=a^2\,n(n+\gamma)
\ee
Since $n\in\mathbb{Z}$ and $\gamma<1$ we have $Q_p\ge 0$ and thus
\be
\eta\equiv {Q_1 Q_5\over Q_1 Q_5 + (Q_1+Q_5) Q_p}\le {Q_1\over Q_p}
\label{in1}
\ee
and
\be
\eta\le 1
\label{in2}
\ee
Let us look at $f+Q_1$ and distinguish the two cases $n>0$ and $n<0$.
(It will be very important here that
$n$ is an integer and that $0<\gamma<1$.) If
$n>0$
\be
f+Q_1\ge Q_1-a^2\,n\,\eta\,\gamma \ge Q_1\,\Bigl(1-{a^2\,n\,\gamma\over
Q_p}\Bigr)=Q_1\,
\Bigl(1-{\gamma\over n+\gamma}\Bigr)>0
\ee
If $n<0$ (and thus $n+\gamma<0$)
\be
f+Q_1\ge Q_1+a^2\,(n+\gamma)\,\eta\,\gamma \ge
Q_1\,\Bigl(1+{a^2\,(n+\gamma)\,\gamma\over Q_p}\Bigr)=
Q_1\,\Bigl(1+{\gamma\over n}\Bigr)>0
\ee
In both cases we used (\ref{in1}). The symmetry of the metric under
interchange of $Q_1$ and
$Q_5$ then also implies $f+Q_5>0$.  Look now at $f+Q_p$. If $n>0$
\be
f+Q_p\ge Q_p - a^2\,n\,\eta\,\gamma = a^2\,n\,(n+(1-\eta)\,\gamma)>0
\ee
as $1-\eta>0$. If $n<0$ (and thus $n+\gamma<0$)
\be
f+Q_p\ge Q_p + a^2\,(n+\gamma)\,\eta\,\gamma = a^2\,(n+\gamma)\,(n+\eta
\gamma)>0
\ee
as $n+\eta \gamma\le n+\gamma<0$. In order to prove that the same
results hold
for the metrics after S,T dualities it is enough to notice that $f$ is
duality
invariant, since the transformation of $\tg_1$ and $\tg_2$ (\ref{sts})
cancel
that of $\eta$:
\be
\eta'= {Q_p Q_5\over  Q_1 Q_5 + (Q_1+Q_5) Q_p}={Q_p\over Q_1}\,\eta
\ee

One can also verify that the metrics (\ref{extremalmetric}) do not have
any horizon. For this purpose we again refer to an argument given in
\cite{gms}: there is no horizon if one can find a timelike vector in the
forward light cone which has a nonzero positive component along the
radial
direction. In turn, the existence of such a vector follows from the fact
that the determinant of the metric restricted to the  $t, y, \psi, \phi$
coordinates
\be
\hat{g}=
-r^2\,(r^2+\eta\,(\tg_1+\tg_2)^2)\,\sin^2\theta\,\cos^2\theta
\ee
is always negative (apart from the points $r=0$ and $\theta=0,\pi/2$,
where
we know there is no horizon by direct analysis\footnote{The naive
geometry with no rotation \cite{bmpv}
has $\tilde\gamma_1+\tilde\gamma_2=0$, and this case has a horizon at
$r=0$; for the  geometries we constructed as duals of actual
microstates $\tilde\gamma_1+\tilde\gamma_2\ne 0$ and the analysis
around $r=0$ done in section 4.2 shows that there is no horizon at
$r=0$.}).

The form of the metric around $r=0$ given in (\ref{genericry}) shows
that the metrics (\ref{genericem}) generically have orbifold 
singularities.
If $R$
is the radius of the $y$ circle, the conical defect parameter is given
by
\be
\gamma= |c_y|\,R
\ee

\section{Inverse metric}\label{app3}
\renewcommand{\theequation}{C.\arabic{equation}}
\setcounter{equation}{0}

The determinant of the extremal metric (\ref{extremalmetric}) is
\be
\sqrt{-g} = h f\,\sin\,\theta\,\cos\,\theta\,r
\label{det}
\ee
The inverse of the metric (\ref{extremalmetric}) is
\bea
&&g^{tt}=-{1\over h f}\Bigl(f+ Q_1+Q_5+Q_p+{Q_1 Q_5 + Q_1 Q_p + Q_5
Q_p\over
r^2+(\tg_1+\tg_2)^2\eta}\Bigr)\nonumber\\
&&g^{yy}={1\over h f}\Bigl(f+ Q_1+Q_5-Q_p+{Q_1 Q_5\,\eta\over r^2}-
{Q_p^2\,\eta\over r^2+(\tg_1+\tg_2)^2\eta}{(Q_1+Q_5)^2\over Q_1
Q_5}
\Bigr)\nonumber\\
&&g^{ty}=-{Q_p\over h f}\Bigl(1+{Q_1+Q_5\over
r^2+(\tg_1+\tg_2)^2\eta }
\Bigr)\nonumber\\
&&g^{\psi\psi}={1\over h f}\Bigl({1\over \cos^2\theta}+{\tg_2^2\,
\eta\over r^2}-{\tg_1^2\,\eta\over
r^2+(\tg_1+\tg_2)^2\eta}\Bigr)
\nonumber\\
&&g^{\phi\phi}={1\over h f}\Bigl({1\over \sin^2\theta}+{\tg_1^2\,
\eta\over r^2}-{\tg_2^2\,\eta\over
r^2+(\tg_1+\tg_2)^2\eta}\Bigr)
\nonumber\\
&&g^{\psi\phi}=-{Q_p\,\eta\over h f}\Bigl({1\over
r^2}-{1\over r^2+(\tg_1+\tg_2)^2\eta}\Bigr)\nonumber\\
&&g^{t\psi}=-{\sqrt{Q_1 Q_5}\over h f}{\tg_1\over
r^2+(\tg_1+\tg_2)^2\eta}\nonumber\\
&&g^{t\phi}=-{\sqrt{Q_1 Q_5}\over h f}{\tg_2\over
r^2+(\tg_1+\tg_2)^2\eta}\nonumber\\
&&g^{y\psi}={\sqrt{Q_1 Q_5} \tg_2\,\eta\over h f}\Bigl({1\over r^2}+
{\tg_1^2\over
r^2+(\tg_1+\tg_2)^2\eta}{Q_1+Q_5\over Q_1 Q_5}\Bigr)\nonumber\\
&&g^{y\phi}={\sqrt{Q_1 Q_5} \tg_1\,\eta\over h f}\Bigl({1\over r^2}+
{\tg_2^2\over
r^2+(\tg_1+\tg_2)^2\eta}{Q_1+Q_5\over Q_1 Q_5}\Bigr)\nonumber\\
&&g^{rr}={r^2 + (\tg_1+\tg_2)^2\eta\over h f}\,,\quad
g^{\theta\theta}={1\over h f}\,,\quad
g^{x_i x_j}= \sqrt{H_5\over H_1}\,\delta^{ij}
\eea

\section{Solution of the wave equation}\label{app4}

\renewcommand{\theequation}{D.\arabic{equation}}
\setcounter{equation}{0}

\subsection{Matching region}
In the matching region $1\ll x\ll \sigma^2$ the wave equation becomes
\be
4{d\over d\,x}\Bigl(x^2{d\over d\,x}\Bigr) H_\mathrm{match} +
(1-\nu^2)\, H_\mathrm{match} = 0
\ee
A basis of independent solutions is
\be
H_\mathrm{match}^{(1)}=x^{\nu-1\over 2}\,,\quad H_\mathrm{match}^{(2)} =
x^{-{\nu+1\over 2}}
\ee

\subsection{Outer region}
In the outer region $x\gg1$ the radial equation is
\be
4{d\over d x}\Bigl(x^2 {d\,H_{\mathrm{out}}\over dx}\Bigr)+
\Bigl[\sigma^{-2}\,x + 1-\nu^2\Bigr]\,H_{\mathrm{out}}=0
\label{outereq}
\ee
with solution
\be
H_\mathrm{out}={1\over\sqrt{x}}\bigl[C_1\,J_\nu(\sigma^{-1}\sqrt{x})+
C_2\,
J_{-\nu}(\sigma^{-1}\sqrt{x})\bigr]
\label{poneapp}
\ee
In the matching region this gives
\bea
H_\mathrm{out}=C_1{1\over\sqrt{x}}\,\Bigl({\sigma^{-2}\,x\over
4}\Bigr)^{{\nu\over 2}}\,
{1\over \Gamma(\nu+1)}\,+C_2 {1\over\sqrt{x}}\,
\Bigl({\sigma^{-2}\,x\over 4}\Bigr)^{-{\nu\over 2}}\, {1\over
\Gamma(-\nu+1)}\,
\label{outer}
\eea

\subsection{Inner region}
In the inner region, $\sigma^{-2}\,x\ll 1$, the wave equation
simplifies to
\be
4{d\over d\,x}\Bigl(x (x + \delta^2){d\over d\,x}\Bigr) H_\mathrm{in} +
\Bigl[1-\nu^2+{\xi^2\over x+\delta^2} -{\zeta^2 \over x} \Bigr]\,
H_\mathrm{in} = 0
\ee
By defining
\be
H_\mathrm{in}=x^\alpha\,(x+\delta^2)^\beta\,F
\ee
with
\be
\alpha={|\zeta|\over 2\,\delta}\,,\quad\beta={\xi\over 2\,\delta}
\ee
the equation above reduces to an hypergeometric equation for $F$. Of
the two
independent solutions, only one is regular at $x=0$:
\be
H_\mathrm{in}= x^\alpha\,(x+\delta^2)^\beta\,
F\Bigl(p,q;1+2\alpha;-{x\over \delta^2}\Bigr)
\ee
where
\be
p={1\over 2}+\alpha+\beta+{\nu\over 2}\,,\quad
q={1\over 2}+\alpha+\beta-{\nu\over 2}
\ee
To get the large $x$ behavior we write
\bea
&&\!\!\!\!\!\!\!\!\!\!\!\!F\Bigl(p,q;1+2\alpha,-{x\over \delta^2}\Bigr)=
{\Gamma(1+2\alpha) \Gamma(-\nu)\over
\Gamma(q) \Gamma({1\over2}+\alpha-\beta-{\nu\over 2})}
\Bigl({x\over \delta^2}\Bigr)^{-p}
F\Bigl(p,p-2\alpha;\nu+1;-{\delta^2\over x}
\Bigr)\nonumber\\
&&\qquad +{\Gamma(1+2\alpha) \Gamma(\nu)\over
\Gamma(p) \Gamma({1\over2}+\alpha-\beta+{\nu\over 2})}
\Bigl({x\over \delta^2}\Bigr)^{-q}
F\Bigl(q,q-2\alpha;-\nu+1;-{\delta^2\over x}
\Bigr)
\eea
and note that
\be
F(p,q;\nu;z)=\sum_{n=0}^\infty {\Gamma(p+n) \Gamma(q+n)\over
\Gamma(p) \Gamma(q)}
{\Gamma(\nu)\over \Gamma(\nu+n)} {z^n\over n!}
\ee
One thus gets (dropping an overall constant)
\bea
&&\!\!\!\!\!
H_\mathrm{in}={1\over\sqrt{x}}\,\Bigl[
{\Gamma(\nu)\over \Gamma(p)\Gamma({1\over 2}+\alpha-\beta+{\nu\over
2})}\,
\Bigl({x\over \delta^2}\Bigr)^{\nu\over 2}\,(1+O(\delta^2
x^{-1}))\nonumber\\
&&\qquad\quad+{\Gamma(-\nu)\over \Gamma(q)
\Gamma({1\over 2}+\alpha-\beta-{\nu\over 2})}\,
\Bigl({x\over \delta^2} \Bigr)^{-{\nu\over 2}}\,
(1+ O(\delta^2 x^{-1}))\Bigr]
\label{inner}
\eea

\subsection{Matching the solutions}

Matching the coefficients of $x^{(-1 \pm \nu)/2}$ between
$H_\mathrm{out}$ and
$H_\mathrm{in}$ we find the ratio $C_2/C_1$ given in (\ref{result}).

\section{Computation of travel time and absorption
probability}\label{app5}
\renewcommand{\theequation}{E.\arabic{equation}}
\setcounter{equation}{0}

In order to write the reflection amplitude $\mathcal{R}$ given in
(\ref{result}-\ref{reflection}) as in (\ref{form}) it is important to
know
the sign  of the arguments
\be
\alpha \pm \beta + {\nu+1\over 2}
\ee
of the Gamma functions appearing in (\ref{result}). As we chose
$\beta\ge 0$,
$\alpha + \beta + {\nu+1\over 2}$ is always positive and thus we are
left with only two
possibilities.

\smallskip

{\bf Case 1:}
\be
\alpha - \beta + {\nu+1\over 2}>0
\label{case1}
\ee
In this case one can use the series expansion
\be
{\Gamma(z+a)\over \Gamma(z+b)}=z^{a-b}\sum_{k=0}^\infty
{(-1)^k\,(b-a)_k\,B(k,a-b+1,a)\,z^{-k}\over k!}
\label{gammaseries}
\ee
($(a)_k$ is the Pochhammer symbol and $B(k,a,b)$ the Bernoulli
polynomial) which is valid for $z+a>0$,
to show that $\mathcal{R}$ has no oscillating factor like
$\exp(2\pi i n \omega/R  \Delta t)$,
and thus there is no absorption.

\smallskip

{\bf Case 2:}
\be
\alpha - \beta + {\nu+1\over 2}<0
\ee

One can rewrite the ratio of Gamma functions in (\ref{result}) with
negative
arguments as:
\bea
&&\!\!\!\!\!\!\!\!\!\!\!\!\!\!\!\!{\Gamma({1\over 2}+
\alpha-\beta+{\nu\over 2})\over
\Gamma({1\over 2}+ \alpha-\beta-{\nu\over 2})}\\
&& \quad = {\Gamma({1\over 2}+
\beta-\alpha+{\nu\over 2})\over
\Gamma({1\over 2}+ \beta-\alpha-{\nu\over 2})}\,
{\sin\Bigl(\pi(\alpha-\beta+{1-\nu\over 2})\Bigr) \over
\sin\Bigl(\pi(\alpha-\beta+{1+\nu\over 2})\Bigr) }\nonumber\\
&&\quad = {\Gamma({1\over 2}+
\beta-\alpha+{\nu\over 2})\over \Gamma({1\over 2}+
\beta-\alpha-{\nu\over 2})}
\,\Bigl[e^{-i\pi\nu}+(e^{-i\pi\nu}-e^{i\pi\nu})\,\sum_{n=1}^\infty
e^{2\pi i\,n (\beta-\alpha -{1+\nu\over 2})}\Bigr]\nonumber
\label{mani}
\eea
so that $\mathcal{R}$ is brought into the form (\ref{form}):
\bea
&&\!\!\!\!\!\!\!\!\!\!\!\!\mathcal{R} = e^{-i\pi\nu}+ {4\pi^2\over
\Gamma^2(\nu) \Gamma^2(\nu+1)}{e^{-i\pi\nu}\over e^{2\pi i\nu}-1}
\Bigl({\delta\over 2 \sigma}\Bigr)^{\!2\nu}{\Gamma({1\over 2}+
\alpha+\beta+{\nu\over 2}) \Gamma({1\over 2}+
\beta-\alpha+{\nu\over 2})\over \Gamma({1\over 2}+
\alpha+\beta-{\nu\over 2})
\Gamma({1\over 2}+ \beta-\alpha-{\nu\over2})}\nonumber\\
&&\!\!\!\!\!\!\!\!\!\!\!\! - {4\pi^2\,e^{-i \pi\nu}\over
\Gamma^2(\nu) \Gamma^2(\nu+1)} \Bigl({\delta\over 2
\sigma}\Bigr)^{\!2\nu}
{\Gamma({1\over 2}+
\alpha+\beta+{\nu\over 2}) \Gamma({1\over 2}+
\beta-\alpha+{\nu\over 2})\over \Gamma({1\over 2}+
\alpha+\beta-{\nu\over 2})
\Gamma({1\over 2}+ \beta-\alpha-{\nu\over 2})}
\sum_{n=1}^\infty e^{2\pi i\,n (\beta-\alpha -{1+\nu\over
2})}\nonumber\\
\label{refapp}
\eea

    From the equation above we can read off the probability of
absorption/emission
(\ref{abs1}) and the time of travel (\ref{time}).

\vfill\eject

\end{document}